\documentclass[conference]{IEEEtran}
\usepackage{fancyhdr}

\usepackage{graphicx}
\usepackage{xcolor}
\usepackage{amsmath}
\usepackage[printonlyused,withpage]{acronym}
\usepackage{algpseudocode,algorithm,algorithmicx}
\usepackage{clrs+}
\usepackage{xspace}
\usepackage{enumitem}
\usepackage{soul}
\usepackage{booktabs}
\usepackage{multirow}
\usepackage{hyperref}

\newcommand{\mA}{\mathbf{A}}
\newcommand{\mB}{\mathbf{B}}
\newcommand{\mC}{\mathbf{C}}
\newcommand{\mS}{\mathbf{S}}
\newcommand{\mE}{\mathbf{E}}
\newcommand{\transpose}     {^{\mbox{\scriptsize \sf T}}}
\newcommand{\kmer}[1][]{\textit{k}-mer#1}

\begin{document}

\renewcommand{\thefootnote}{\fnsymbol{footnote}}
\title{Distributed Many-to-Many Protein Sequence Alignment using Sparse Matrices}
\author{
  \IEEEauthorblockN{
    Oguz Selvitopi\IEEEauthorrefmark{1}\textsuperscript{1},
    Saliya Ekanayake\IEEEauthorrefmark{2}\textsuperscript{1},
    Giulia Guidi\IEEEauthorrefmark{3}\IEEEauthorrefmark{1},
    Georgios A. Pavlopoulos\IEEEauthorrefmark{4},
    Ariful Azad\IEEEauthorrefmark{5},
    Ayd{\i}n Bulu\c{c}\IEEEauthorrefmark{1}\IEEEauthorrefmark{3}}
  \vspace{0.5em}
  \IEEEauthorblockA{
    \IEEEauthorrefmark{1}\textit{Computational Research Division, Lawrence Berkeley National Laboratory, USA} \\
    \IEEEauthorrefmark{2}\textit{Microsoft Corporation, USA} \\
    \IEEEauthorrefmark{3}\textit{University of California, Berkeley, USA} \\
    \IEEEauthorrefmark{4}\textit{Institute for Fundamental Biomedical Research, BSRC ``Alexander Fleming'', 34 Fleming Street, 16672, Vari, Greece} \\
    \IEEEauthorrefmark{5}\textit{Indiana University, USA} \\
    roselvitopi@lbl.gov}   
  }

\maketitle
\begingroup\renewcommand\thefootnote{1}
\footnotetext{Equal contribution.}
\thispagestyle{fancy}
\lhead{}
\rhead{}
\chead{}
\lfoot{\footnotesize{
    SC20, November 9-19, 2020, Is Everywhere We Are
    \newline 978-1-7281-9998-6/20/\$31.00 \copyright 2020 IEEE}}
\rfoot{}
\cfoot{}
\renewcommand{\headrulewidth}{0pt}
\renewcommand{\footrulewidth}{0pt}

\newcommand{\swname}{PASTIS}

\acrodef{BLAST}{Basic Local Alignment Search Tool}
\acrodef{PSI-BLAST}{Position Specific Iterated Basic Local Alignment Search Tool}
\acrodef{MMseqs2}{Many-against-Many sequence searching}
\acrodef{\swname}{Protein Alignment via Sparse Matrices}
\acrodef{PSG}{Protein Similarity Graph}
\acrodef{MPI}{Message Passing Interface}
\acrodef{GPU}{Graphics Processing Unit}
\acrodef{CombBLAS}{Combinatorial BLAS}
\acrodef{SW}{Smith-Waterman}
\acrodef{XD}{seed-and-extend with x-drop}
\acrodef{FASTA}{FAST-All}
\acrodef{BELLA}{Berkeley Efficient Long-Read to Long-Read Aligner and Overlapper}
\acrodef{diBELLA}{Distributed \ac{BELLA}}
\acrodef{SpGEMM}{Sparse General Matrix Multiply}
\acrodef{SeqAn}{The Library for Sequence Analysis}
\acrodef{AVX2}{Advanced Vector Extensions 2}
\acrodef{HipMCL}{High-performance Markov Clustering}
\acrodef{ANI}{Average Nucleotide Identity}
\acrodef{NS}{Normalized Raw Alignment Score}
\acrodef{NW}{Needleman-Wunsch}
\acused{LAST}
\acused{CPU}
\acrodef{DIAMOND}{Double Index Alignment of Next-generation Sequencing Data}

\begin{abstract}
  \small\baselineskip=9pt
  Identifying similar protein sequences is a core step
  in many computational biology pipelines such as detection of homologous
  protein sequences, generation of similarity protein graphs for downstream
  analysis, functional annotation, and gene location. Performance and
  scalability of protein similarity search have proven to be a bottleneck in
  many bioinformatics pipelines due to increase in cheap and abundant sequencing
  data. This work presents a new distributed-memory software
  PASTIS. PASTIS relies on sparse matrix computations for efficient identification
  of possibly similar proteins. We use distributed sparse matrices for
  scalability and show that the sparse matrix infrastructure is a great fit for
  protein similarity search when coupled with a fully-distributed dictionary of
  sequences that allow remote sequence requests to be fulfilled. Our algorithm
  incorporates the unique bias in amino acid sequence substitution in search
  without altering basic sparse matrix model, and in turn, achieves ideal
  scaling up to millions of protein sequences.

\end{abstract}

\section{Introduction} \label{sec:introduction}
One of the most fundamental tasks in computational biology is \emph{similarity search}.
Its variants can be used to map short DNA sequences to a reference genome or find homologous regions between nucleotide sequences, such as genes that are the basic functional unit of heredity. 
A pair of genes is homologous if they both descend from a common ancestor. 
The DNA sequence of a gene contains the information to build amino acid sequences constituting proteins. 

Computing sequence similarity is often used to infer homology because homologous sequences share significant similarities despite mutations that happen since the evolutionary split.
While the similarity computation can be carried out in the DNA sequence space, it is more often performed in the amino acid sequence space.
This is because the amino acid sequence is less redundant, resulting in fewer false positives.
In this paper, we will refer to this problem of inferring homology in the amino acid space as \emph{protein homology search}.

Protein homology search has numerous applications.
For example, functional annotation uses known amino acid sequences to assign
functions to unknown proteins.
Another example area of application is gene localization, which is crucial to
identify genes that may affect a given disease or gain insight about a
particular functionality of interest.
The primary motivation of our work is the identification of protein families, which are groups of proteins that descend from a common ancestor.
It is a hard problem since the relationship between sequence similarity and homology is imprecise, meaning that one cannot use a similarity threshold to accurately conclude that two proteins are homologous or belong to the same family. 

Different algorithms attempt to infer homology directly via similarity search~\cite{li2006cd, edgar2010search} with variable degree of success in terms of sensitivity and specificity.
Alternatively, one can perform a similarity search within a protein data set and  construct a similarity graph~\cite{steinegger2017mmseqs2, Altschul1997, pmid:21209072, buchfink2015fast}, and then feed it into a clustering algorithm~\cite{enright2002efficient, wittkop2010partitioning, azad2018hipmcl, Ruan:2012:DDA:2382936.2382978} to ultimately identify protein families.
The advantage of the latter approach is that the clustering algorithm can use global information to determine families more accurately.
For example, if two proteins $P1$ and $P2$ are incorrectly labeled as homologous by the similarity search, this false positive link can be ignored by the clustering algorithm if it is not supported by the rest of the similarity graph (say if the neighbors of $P1$ and $P2$ are distinct except for $P1$ and $P2$ themselves).
Likewise, missed links (false negatives) can be recovered by the clustering algorithm using the topology of the graph.
The disadvantage of this approach lies in the computational cost of storing the similarity graph and executing of the subsequent clustering algorithm.
Dual-purpose tools such as \ac{MMseqs2}~\cite{steinegger2017mmseqs2} can be used to either directly cluster proteins or to generate a protein similarity graph, depending on their settings. 

Recent advancements in shotgun metagenomics, where a metagenome is collected from an environmental sample and it is thus composed of DNA from different species, are leading to the expansion of the known protein space~\cite{godzik2011metagenomics}.
The massive amount of data associated with modern isolate genome and metagenome databases have rendered many existing tools too slow in constructing protein similarity graphs.
Distributed parallel computation is one way to efficiently manipulate this data deluge.

In this paper, we present a fully distributed pipeline called \ac{\swname} for
large-scale protein similarity search.
\ac{\swname} constructs similarity graphs from large collections of protein
sequences, which in turn can be used by a graph clustering algorithm to
accurately discover protein families.
A major novelty of \ac{\swname} is its use of distributed sparse matrices as its
underlying data structure.
Not only the sequences and their $k$-mers are stored through sparse matrices,
but also the substitute $k$-mers that are critical for controlling sensitivity
and specificity during sequence overlapping.
We develop custom semirings in sparse matrix computations to enable different
types of alignments.
\ac{\swname} extensively hides communication and exploits the symmetricity of
the similarity matrix to achieve load balance.
We present detailed evaluations about parallel performance and relevance.
We demonstrate the scalability of \ac{\swname} by scaling it up to 2025 nodes
(137,700 cores) and show that its accuracy is on par with \ac{MMseqs2}.

%
%
%

Our resulting software, PASTIS, is available publicly as open source at \href{https://github.com/PASSIONLab/PASTIS}{\color{blue}{https://github.com/PASSIONLab/PASTIS}}.
The rest of this paper is organized as follows.
Section~\ref{sec:background} describes the problem we address and
Section~\ref{sec:related_work} presents the related work.
In Section~\ref{sec:methods} we describe our methodology for forming the
similarity graph and then we focus on its parallel aspects in
Section~\ref{sec:impl}.
We evaluate our approach in Section~\ref{sec:experiments} and we conclude in
Section~\ref{sec:conclusion}.




%

\section{Background} \label{sec:background}
Protein homology search, as introduced in Section~\ref{sec:introduction}, is modeled as a sequence similarity searching problem. 
Herein, we formally define the notation and the problem.

Let $S = \{s_1, s_2,\dots,s_n\}$ be a set of $n$ protein sequences.  Define the
\ac{PSG} graph $G = (V, E)$ as $V=S$ and $E=\{(s_i,s_j)~|~s_i$ and $s_j$ exceed
a given similarity threshold$\}$.  The weight of an edge $(s_i,s_j)$ is denoted
with $w(s_i,s_j)$ and it indicates the strength of similarity between sequences
$s_i$ and $s_j$.  A \emph{\kmer{}} or seed is defined as a subsequence of a
given sequence $s$ with fixed length $k$.

The problem we address in this work is defined as follows.  Given $p$ processing
elements, a set $S$ of proteins, and similarity constraints, compute $G=(V,E)$
efficiently in parallel.
Figure~\ref{fig:homology_detection_pipeline} depicts the base pipeline to form
the similarity graph.

\subsection{\ac{CombBLAS}}
\label{sec:combblas}
\ac{CombBLAS}~\cite{bulucc2011combinatorial} is a distributed memory parallel
graph library that is based on sparse matrix and vector operations on arbitrary user-defined semirings.
A semiring consists of two binary operators, addition and multiplication, that
satisfy certain requirements.
\ac{CombBLAS} allows users to define their own types for matrix and vector
elements and overload the operations on sparse matrices.
This allows it to express a broad range of algorithms that operate on graphs.
%

\ac{CombBLAS} supports MPI/OpenMP hybrid parallelism and uses a 2D block
decomposition for distributed sparse matrices.
%
%
The 2D decomposition of the matrix constrains most of the communication
operations to rows/columns of the process grid and enables better scalability
than the 1D decomposition.
%
Among several operations supported by \ac{CombBLAS}, \ac{SpGEMM} is heavily
utilized by \ac{\swname} and it contains several distinct optimizations.
%
%
Distributed \ac{SpGEMM} in \ac{CombBLAS} uses a scalable 2D Sparse
SUMMA~\cite{Buluc2012} algorithm, and for the local multiplication it uses a
hybrid hash-table and heap-based algorithm that is faster than the existing
libraries~\cite{Nagasaka2019}.
Another distributed matrix library that supports semiring algebra on sparse 
matrices is the Cyclops Tensor Framework (CTF)~\cite{solomonik2015sparse}.
Both CTF and CombBLAS support 2D as well as 3D SpGEMM algorithms. 

\begin{figure}[t]
    \centering
    \includegraphics[width=1\columnwidth]{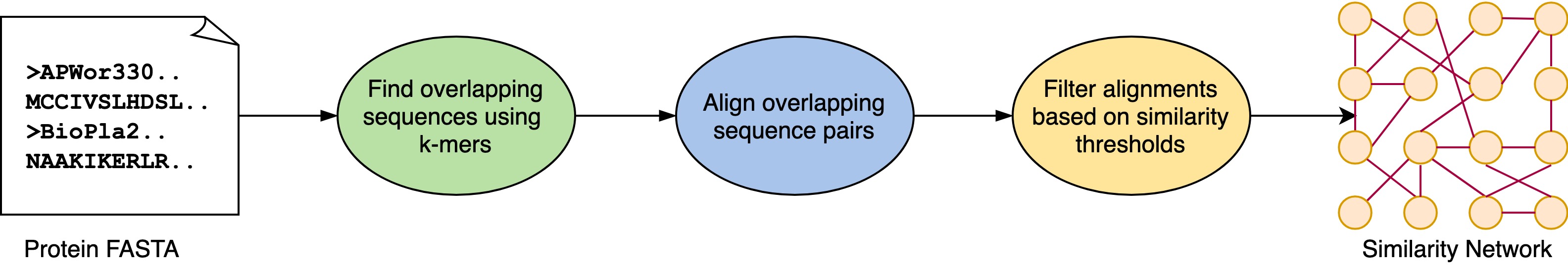}
    \vspace{-1.5em}
    \caption{Homology detection pipeline.}
    \vspace{-1.5em}
    \label{fig:homology_detection_pipeline}
\end{figure}

\section{Related Work} \label{sec:related_work}

pGraph~\cite{wu2012pgraph} is a distributed software to build protein homology
graph.
It is similar to \ac{\swname} but with a different distributed implementation
and a different way of detecting homologous sequences.
Given an input sequence set $S$, it uses a suffix tree based
algorithm~\cite{kalyanaraman2003space} to identify pairs of sequences that pass
a user-defined criteria.
These pairs are then aligned using the Smith-Waterman algorithm~\cite{smith1981identification}.
pGraph distributed implementation includes a super-master process and a collection of subgroups.
Each subgroup consists of a master, a set of producers, and a set of consumers.
The producers in each subgroup are responsible for the generation of sequence pairs in parallel, while the consumers perform alignment on those pairs.
The master regulates producers and consumers such that no overactive producers
exist in the system while at the same time keeping consumers busy.
The super-master acts as a regulatory body across all subgroups.
One caveat with the producer-consumer approach is that sequence data corresponding to a generated pair may not lie within the local memory of the consumer handling that pair.
pGraph introduces two options to overcome this problem: (a) reading from disk or (b) communicating remote sequences from other processes.
Both techniques incur latencies, however, their experiments suggest that remote
sequence fetching over the network is more efficient than reading from disk.
Their results show linear scaling for a set of $\approx2.5$ million sequences, where a
total of $5.3$ billion pairs were aligned.

\ac{BELLA} is a shared-memory software for overlap detection and alignment for long-read \emph{de novo} genome assembly and error correction~\cite{guidi2018bella}, where \emph{long-read} indicates a category of sequencing data.
Despite different objectives, \ac{BELLA} is the first work to formulate the overlap detection problem as a \ac{SpGEMM}.
It uses a seed-based approach to detect overlaps and uses a sparse matrix, $\mA$, to represent its data, where the rows represent nucleotide sequences and columns represent $k$-mers. 
$\mA$ is then multiplied by $\mA\transpose$, yielding a sparse \emph{overlap
  matrix} $\mA \mA\transpose$ of dimensions $S$-by-$S$, where each non-zero cell
$(i,j)$ of the overlap matrix stores the number of common $k$-mers between the
$i$th and $j$th sequences, and their positions in the corresponding sequence
pair. In this work, we adopt this technique and extend it to implement a distributed memory \kmer{} matching.
\ac{BELLA} also has a distributed version, diBELLA~\cite{ellis2019dibella}, that is being developed.
However, it operates on nucleotide sequences and computes overlap detection
using distributed hash tables rather than distributed \ac{SpGEMM}.


\ac{MMseqs2}~\cite{steinegger2017mmseqs2} is a software to find target sequences similar to a given query sequence by searching a precomputed index of target sequences.
A target sequence is chosen to be aligned against the query only if they share two similar \kmer{s} along the same diagonal.
The notion of similar \kmer{s} is comparable to the substitute \kmer{s} of
our work (Section~\ref{subsubsec:methods:find_overlaps_withsubstitutes}).
The authors claim the two-\kmer{} approach increases the sensitivity lowering the probability of that match happening by chance.
Notably, a pair could have more than one diagonal containing two similar \kmer{s}.
Once a pair is chosen, an ungapped alignment is performed on the \kmer{s} on each diagonal.
Additionally, a gapped alignment is performed if the diagonal with the best ungapped score passes a given threshold.
\ac{MMseqs2} is a faster and more sensitive search tool compared to its popular counterpart \ac{BLAST} and it can run in parallel on distributed memory systems.

LAST~\cite{pmid:21209072} is another heuristics-based sequence search tool.
LAST also uses matching subsequences to identifying similar sequences but it supports a richer notion of seed than the regular \kmer{} one.
LAST provides both spaced seeds and substitute seeds.
A spaced seed can be thought of as supporting the wildcard `*' in the seed definition, so that certain positions of a \kmer{} could be ignored or matched regardless of the character (or amino acid) in the target sequence at that position.
A substitute seed modifies this idea by restricting the wildcard matching to a group of designated characters at each position.
Besides introducing extended seed patterns, LAST takes a step further allowing adaptive seed length.
This features allows LAST to increase the match sensitivity by repeatedly matching a seed pattern until the number of matches in the target sequence matches or drops below a frequency threshold~\cite{pmid:21209072}.
Currently, LAST implementation is based on suffix array and runs in a single compute node with optional shared memory parallelism.
Despite the sensitivity advantages, the shared memory parallelism is a bottleneck for large data sets as its runtime is in the order of days.

\begin{figure}
    \centering
    \includegraphics[width=1\columnwidth]{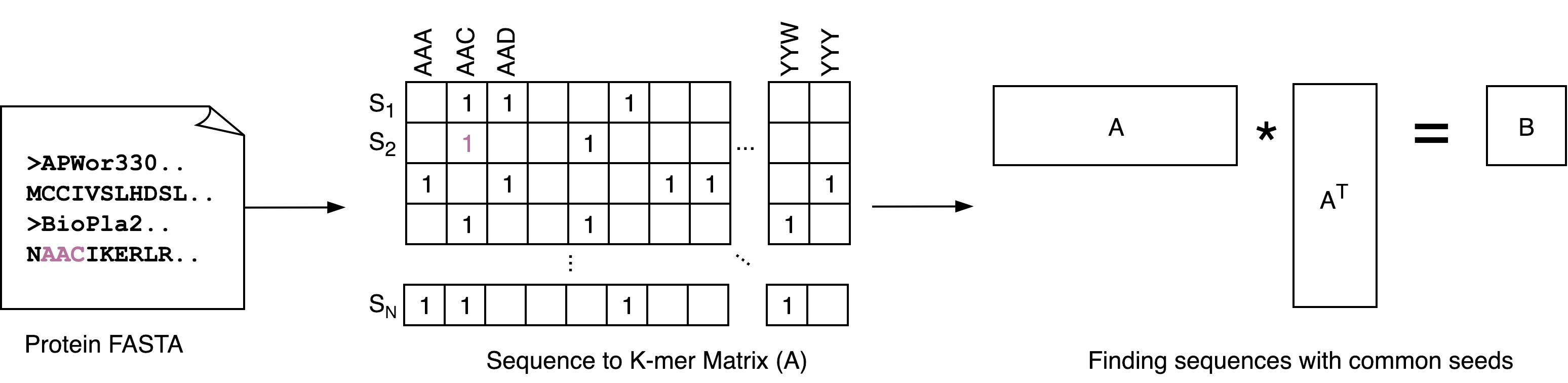}
    \caption{Overlap detection in \ac{\swname}.}
    \label{fig:pisa_find_overlaps}
    \vspace{-1.5em}
\end{figure}

\ac{DIAMOND}~\cite{buchfink2015fast} utilizes a double indexing approach which
determines the list of all seeds in both reference and query sequences.
The double indexing strategy is cache-friendly as it increases the data
locality.
To attain high sensitivity, \ac{DIAMOND} utilizes spaced seeds of certain weight
and shape.
Different shape and weight combinations can be used to achieve different levels
of sensitivity.
Another important feature of \ac{DIAMOND} is that it uses a reduced amino acid
alphabet for greater sensitivity.
This also makes \ac{DIAMOND} more memory-efficient as it results in smaller
index sizes.

\section{\ac{\swname} Concepts} \label{sec:methods}


The \ac{PSG}, $G=(V,E)$, can technically be a clique, where an alignment algorithm would perform $O(n^2)$ comparisons and find some similarity between each pair of sequences. 
Nevertheless, only pairs above a certain similarity threshold are assumed to be biologically related via a common ancestor.
This enables \ac{\swname} to avoid the expensive $O(n^2)$ alignments and exploit efficient sparse matrix operations in constructing the \ac{PSG}.
\ac{\swname} adheres to the pipeline shown in Figure~\ref{fig:homology_detection_pipeline} to compute the \ac{PSG}.
Other heuristics-based sequence alignment tools, described in Section~\ref{sec:related_work}, also follow similar stages.
In the following sections, we describe how these stages are handled in \ac{\swname}.

\begin{figure}[t]
  \centering
  \vspace{-1em}
    \includegraphics[width=0.5\columnwidth]{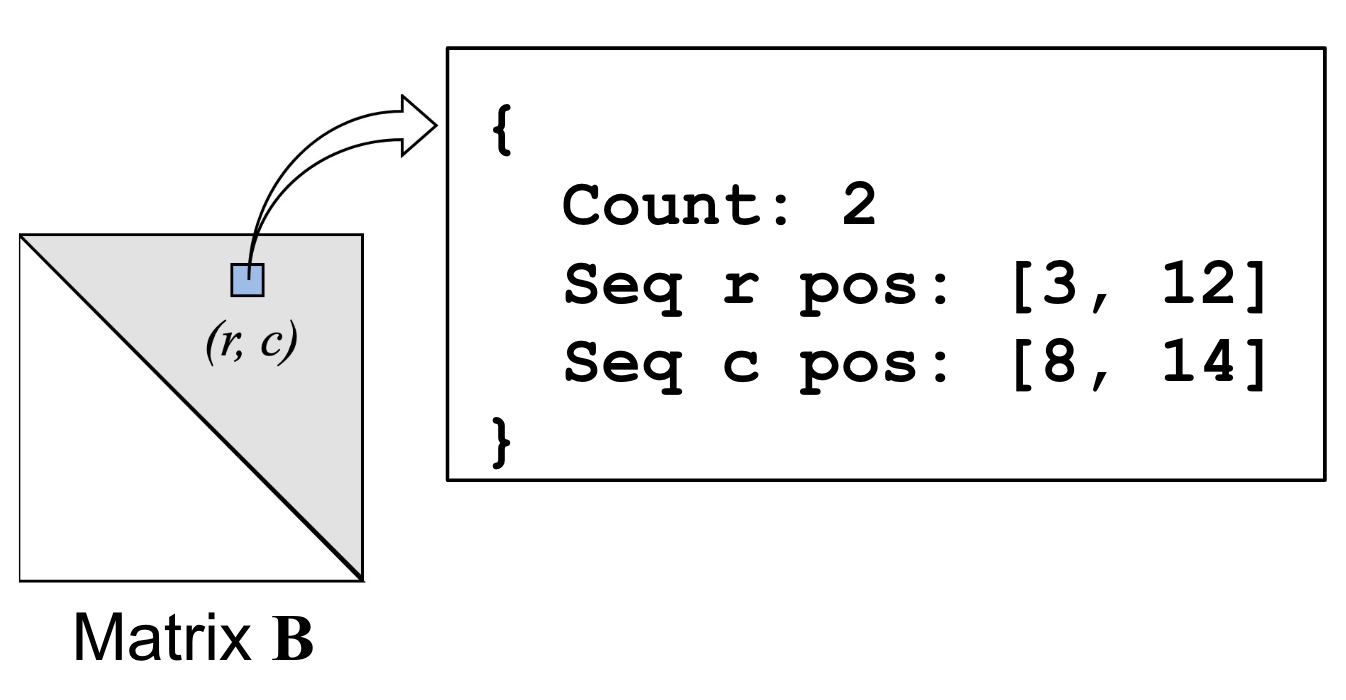}
    \vspace{-1em}
    \caption{The structure of matrix $\mB$ in \ac{\swname}.}
    \vspace{-1em}
    \label{fig:matrix_B}
\end{figure}

\begin{figure*}[t]
    \centering
    \begin{minipage}{.54\textwidth}
        \centering
        \includegraphics[width=0.7\columnwidth]{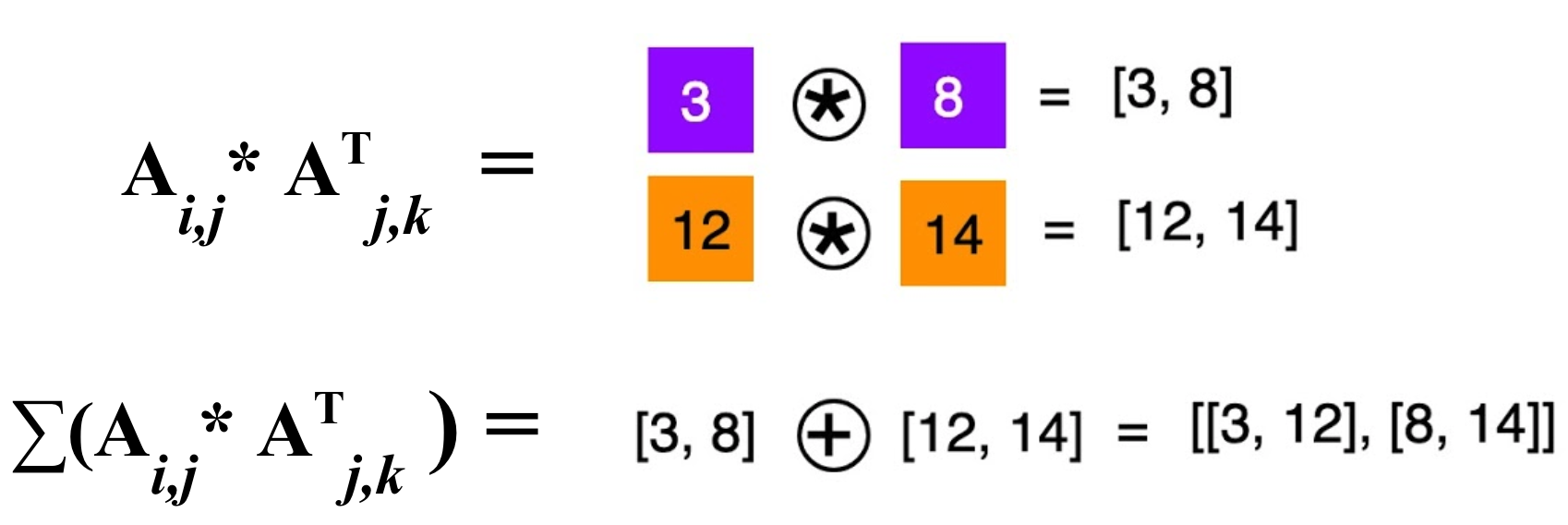}
    \end{minipage}
    \begin{minipage}{0.35\textwidth}
        \centering
        \includegraphics[width=0.7\columnwidth]{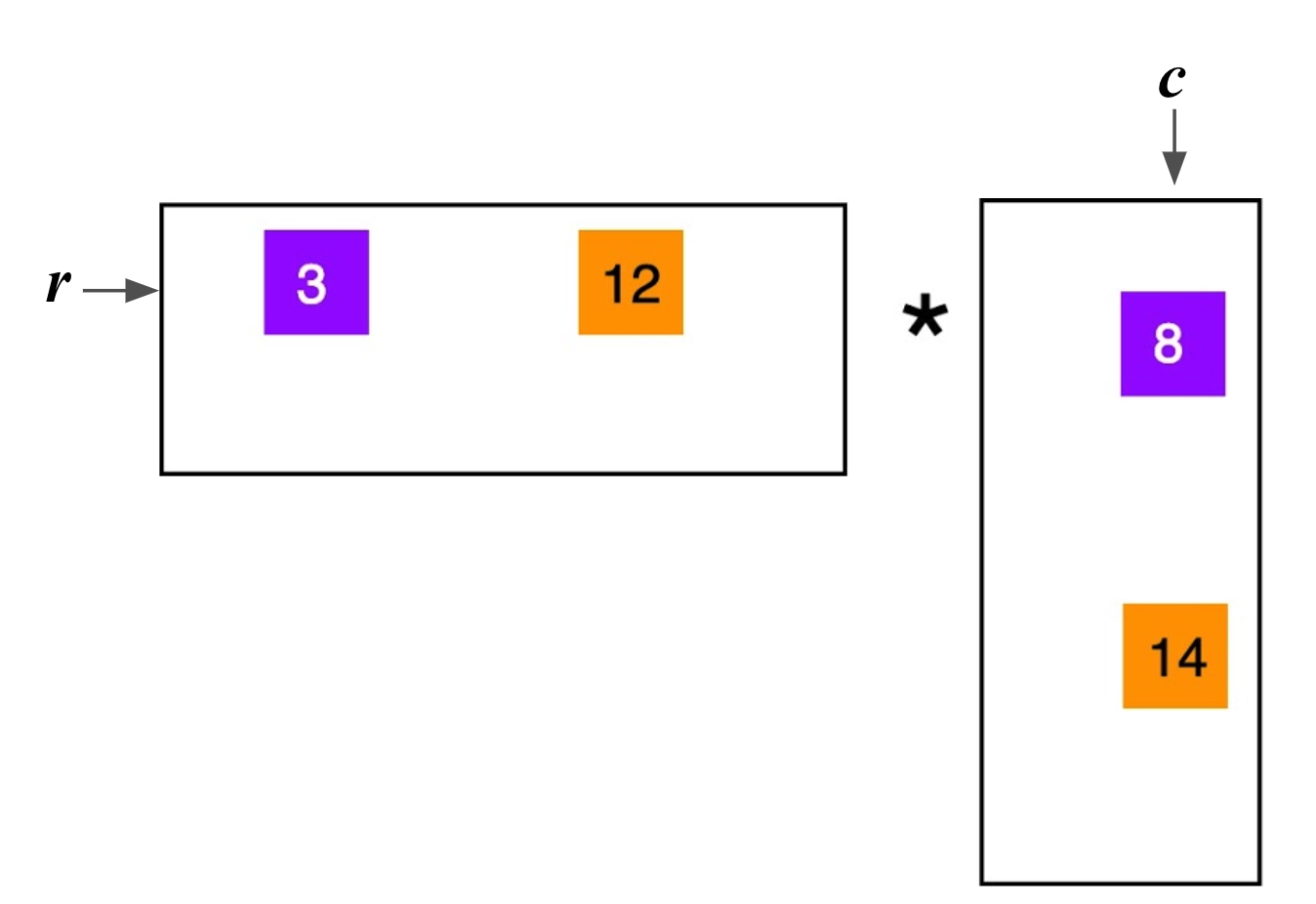}
    \end{minipage}
    \vspace{-1.5em}
    \caption{Semiring for exact \kmer{} matching.}
    
    \label{fig:pisa_semiring}
\end{figure*}

\subsection{Overlapping Sequences}\label{subsec:methods:find_overlaps}

\ac{\swname} uses the presence of common \kmer{s} as a heuristic metric, which
is a common technique that is also adopted in tools such as
BLAST~\cite{altschul1990basic}, LAST~\cite{pmid:21209072}, and
\ac{MMseqs2}~\cite{steinegger2017mmseqs2}.
One approach to find these common $k$-mers is to create an index, such as a suffix array, of the input target sequence set and query the same sequence against it.
A distributed implementation of this method would require a parallel implementation of the suffix array.
Alternatively, we use a sparse matrix based implementation that is highly parallel using existing libraries such as \ac{CombBLAS}~\cite{bulucc2011combinatorial}.

Figure~\ref{fig:pisa_find_overlaps} shows the creation of the
$\lvert\textit{sequences}\rvert$-by-$\lvert\textit\kmer{s}\rvert$ matrix $\mA$.
A nonzero entry $\mA_{ij}$ denotes the existence of \kmer{} $j$ in sequence
$i$.
The alphabet determines the number of all possible \kmer{s}. %
For proteins, there are a total of 24 amino acid bases making the size
of the \kmer{} space $24^k$.
Once this matrix is constructed, we compute $\mB = \mA \mA\transpose$ of size $n
\times n$.
If each nonzero entry in $\mA$ was marked as $1$ then $\mB_{ij}$ would perform
exact \kmer{} matching and be equal to the count of common \kmer{s} between
sequences $i$ and $j$.
Consequently, we proceed to align sequence pairs $(i,j)$ where $\mB_{ij}\ne0$.

Using \ac{SpGEMM} to find sequences that share at least one \kmer{} has been
originally proposed in \ac{BELLA}~\cite{guidi2018bella} in the context of
overlapping of long error-prone sequences.  Importantly, \ac{BELLA} work did not
use distributed data structures nor did it use the substitute \kmer{s} concept
we discuss in the next section.

\begin{figure}[h]
    \centering
    \includegraphics[width=1.0\columnwidth]{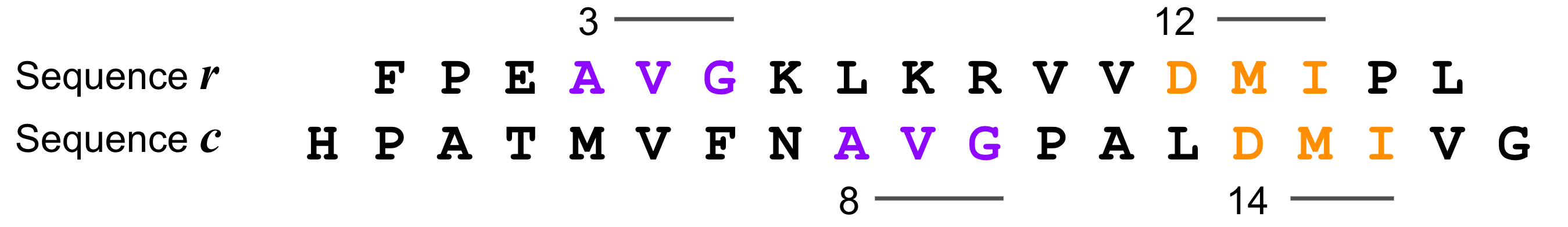}
    \vspace{-1em}
    \caption{Common \kmer{s} of two sequences.}
    \vspace{-1.0em}
    \label{fig:matrix_B_sequences}
\end{figure}

\ac{\swname} stores the positions of the shared \kmer{s} in each sequence along
with the k-mer count as such information is useful in the alignment step.
For example, Figure~\ref{fig:matrix_B_sequences} shows the location of two
common \kmer{s}, $\mathtt{AVG}$ and $\mathtt{DMI}$, on a pair of sequences $r$
and $c$.
Figure~\ref{fig:matrix_B} shows how this information is recorded in $\mB_{rc}$.
Notably, $\mB$ is symmetric.
Therefore, we process only nonzeros belonging to strictly lower or upper
triangular portion of it.

In order to yield the custom structure in $\mB$ as shown above, we employ a
custom semiring to overload the addition and multiplication operators in $\mA
\mA\transpose$.
In addition, $\mA_{ij}$ represents the starting position of \kmer{} $j$ in
sequence $i$ instead of $1$ for its presence.
Figure~\ref{fig:pisa_semiring} shows the matrix multiplication with a custom
semiring in \ac{\swname} when exact \kmer{} matching is used.
%
In \ac{\swname}, the multiplication operator saves the position on the two
sequences for \kmer{} $j$ into a pair while the addition operation organizes the
seed positions on sequence $r$ into one list and positions of $c$ into another
list.
Currently, a maximum of two shared \kmer{} locations per sequence pair are kept
out of all such possible pairs.
While \kmer{s} attempt to capture the latent features of protein sequences,
there is no accepted standard as to which \kmer{s} yield better results than
others.
In future, we plan to study the effect of different number of \kmer{s} as well
as the distance between \kmer{s}.


\subsection{Substitute K-mers} \label{subsubsection:methods:find_overlaps:subkmers}

$\mA \mA\transpose$ results in exact \kmer{} matches.
However, our experiments suggest that exact matches pose an excessively strict
constraint on the overlapping landscape yielding to significantly low recall.
In this context, recall is defined as the ratio between sequence pairs belonging to the same family in both \ac{\swname} output and the original data.
Herein, we introduce the notion of \emph{substitute \kmer{s}} as an approach to
improve the recall of this step (see Section~\ref{sec:prec-recall} for results
about accuracy).  Importantly, this is different from the notion of substitute
seeds used in LAST~\cite{pmid:21209072}, which is about specifying a seed as a
pattern similar to a regular expression. The similar \kmer{s} concept of \ac{MMseqs2} is closest to our approach with the difference being \ac{MMseqs2} limiting similar \kmer{} sets through a scoring threshold whereas \ac{\swname} defines a fixed-sized neighborhood.

In the state-of-the-art, scoring matrices such as BLOSUM62~\cite{pmid:1438297} are used to quantify the chance of a certain amino acid being substituted by another during evolution.
While these scores are typically used during the alignment phase, we
adopt the notion of ``evolution'' for \kmer{s} by producing a set of
substitute \kmer{s} for any given \kmer{}.
Given a set of substitute \kmer{s}, the algorithm chooses the one with the highest chance to appear in-place of the original \kmer{}.

\begin{figure}[b]
    \centering
    \includegraphics[width=0.75\columnwidth]{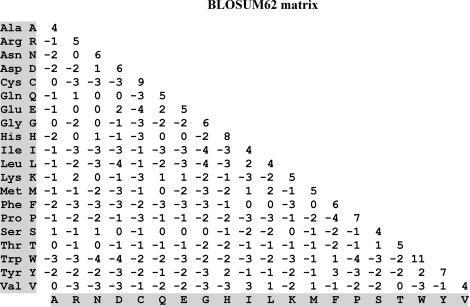}
    \caption{The BLOSUM62 scoring matrix for proteins.}
    \vspace{-1.5em}
    \label{fig:blosum62}
\end{figure}
 
For example, under BLOSUM62 the $\mathrm{3}$-mer $\mathtt{AAC}$ will have a
score of $4+4+9=17$ for an exact match.  As seen in Figure~\ref{fig:blosum62},
the base $A$ can be substituted with $S$ for the least amount of penalty.  Thus,
$\mathrm{3}$-mers $\mathtt{SAC}$ and $\mathtt{ASC}$ both have a score of
$1+4+9=14$ when matched with $\mathtt{AAC}$.  Continuing with this idea, the
next $\mathrm{3}$-mer closest to $\mathtt{AAC}$ is $\mathtt{SSC}$, which has a
match score of $1+1+9=11$.  In \ac{\swname}, we restrict ourselves to computing
$m$ ``nearest'' such substitute \kmer{s} for any original \kmer{} present in a
sequence.

We remark that $m$-nearest \kmer{s} are not restricted to single substitutions.
Depending on the scoring matrix and the values of $m$ and $k$, the $m$-nearest neighbors of a given \kmer{} include \kmer{s} can be multiple hops away in terms of the edit distance.
This can be true even when $m$ is significantly smaller than $k (\lvert
\Sigma \rvert -1)$ where $\Sigma$ is the alphabet.
If we consider our original example $\mathtt{AAC}$, Figure~\ref{fig:blosum62} shows that substitutions in $\mathtt{C}$ cost more than substitutions in $\mathtt{A}$ under BLOSUM62.
Given that a match in $\mathtt{C}$ provides a gain of $9$ in the score, even the least costly substitution lowers the score (e.g., to $\mathtt{M}$ whose score is $-1$) by $10$.
By contrast, a match in $\mathtt{A}$ only has a score of $4$.
Substituting it with $\mathtt{C}$, $\mathtt{D}$, or $\mathtt{T}$, all of which have scores $0$, would only lower the cost by $4$.
Therefore, $3$-mers of the form $\{T | C | G\}\{T | C | G\}C$ are all closer to $\mathtt{AAC}$ (at a ``distance'' of 8) than any $3$-mer of the form
$\mathtt{AA*}$ (except $\mathtt{AAC}$ itself), despite requiring
two letter substitutions.

\begin{algorithm}[t]
  \small
    \caption{Find the nearest $m$ substitute \kmer{s} of a given \kmer{} $r$,
    using sorted expense matrix $\mE$.}
    \label{alg_subk}
    \begin{algorithmic}[1] 
        \Procedure{FindSubKMers}{$r, \mE, m$}
            \State $\id{nbrs} \gets \{\}$ \Comment{neighbor list}
            \State $\id{minmaxheap} \gets \{\}$
            \State $\Call{Explore}{r, \id{minmaxheap}, r, \mE, m}$
            \While{$\lvert \id{nbrs} \rvert < m$}
                \State $\id{mink} \gets \Call{FindMin}{\id{minmaxheap}}$
                \State $\id{nbrs} \gets \id{nbrs} \cup \id{mink}$
                \State $\Call{Explore}{\id{mink}, \id{minmaxheap}, r, \mE, m}$
                \State $\Call{ExtractMin}{\id{minmaxheap}}$
            \EndWhile
            \State \textbf{return} $\id{nbrs}$
        \EndProcedure
    \end{algorithmic}
\end{algorithm}
\setlength{\textfloatsep}{5pt}

Given non-uniform scores, the efficient generation of $m$-nearest \kmer{s} is non-trivial.
For each \kmer{} in the data set, we first generate its 1-hop neighbors (i.e., single-substitution \kmer{s}) that are the $m$ nearest.
This can be done significantly faster than the naive $O(k (\lvert
\Sigma \rvert -1))$ approach if $m$ is small.
For each letter in $\Sigma$, we can pre-compute the lowest cost substitutions by simply sorting (in decreasing values) the off-diagonal entries in the corresponding column (or row because the scoring matrix is symmetric) of the scoring matrix.
This pre-computation only needs to be done once per scoring matrix rather than for each \kmer{}, therefore the cost is minuscule.
Assuming $m < k (\lvert \Sigma \rvert-1)$, we now just need to simply merge
these sorted lists into a single sorted list of length $m$.
The crux is that we do not need to touch the entire set of columns during this merging because we only need the top $m$ elements of the final merged list.
Hence, the cost of initial list generation is $O(m)$.
This initial list is solely composed of single-substitution \kmer{s} and no other single-substitution $k$-mer can be closer to our seed \kmer{}.

From here, we run an algorithm in the spirit of Dijkstra's shortest path algorithm.
The differences are that (1) we are only interested in paths that are up to length $m$, (2) the edges in our case are implicit (i.e., never materialized) and only generated as needed, and (3) this implicit graph is acyclic; in fact, we are exploring a tree with a branching factor of $(\lvert \Sigma \rvert -1)$.
Properties (1) and (3) allow us to stop exploring substitutions in a given position once we know that the current distance is farther than the current $m$-nearest neighbors list.
Those positions are marked as \emph{inactive}.
This current $m$-nearest neighbors list is implemented using a max heap (priority queue).
For each active position, we iterate over substitutions in increasing distance, using the sorted columns of our scoring matrix.
If a substitution gives a score lower than the top of our heap, we push it to our heap.
Note that checking the top of the heap with $\proc{FindMin}$ is $O(1)$ whereas insertion and extraction are $O(\log(m))$, hence it only costs when we find a new $m$-nearest neighbor.
If a substitution does not give a score lower than the top of our heap, we mark that position as \emph{inactive} and not explore any further substitutions.
This is possible because we are exploring substitutions in increasing cost, and thus no other substitution on that position can make it to the $m$-nearest
neighbor list.
The algorithm stops when all of the $k$ \kmer{} positions are marked as \emph{inactive}.

The pseudocode to find the most possible substitute \kmer{s} is described in
Algorithm~\ref{alg_subk}.
Here, the scoring matrix whose row entries are sorted is denoted by $\mE$
because it encodes the ``expense'' we incur in order to substitute the
amino acid bases in those \kmer{s}.
If the substitution matrix is denoted by $\mC$, then $\mE = \proc{Sort}( \proc{Diag}(\mC)- \mC)$, where the function $\proc{Diag()}$ simply creates a diagonal matrix out of its parameter by deleting its off-diagonal entries, and $\proc{Sort()}$ sorts the rows of a matrix in ascending order.

$\mE$ stores both the integer expenses and the bases corresponding to that expense in its auxiliary field, which is accessed by accessible via $\texttt{.base}$.
For example, the first row of $\mE$ would be: $\mE[1] = \{(0,\mathtt{A}), (3,\mathtt{S}), (4,\mathtt{C}), (4,\mathtt{G}), \ldots \}$ for the BLOSUM62 shown in Figure~\ref{fig:blosum62}.
The first entry of each row is somewhat uninteresting and $\mE[i][1]$ (the arrays are 0-indexed) would give the cheapest substitution to the $i$th base.
Each individual base in \kmer{s} can be accessed and modified using
$\texttt{operator[]}$.
Furthermore, each \kmer{} object has a list of ``free'' indices, which are locations that have not been previously substituted by the algorithm.
At the beginning, the entire list of indices $[0, k-1]$ is free.


\begin{algorithm}[t]
  \small
    \caption{Explore the next nearest \kmer{s} of a given \kmer{} $p$, with
    respect to a root \kmer{} $r$ and an existing set of nearest \kmer{s}.
    The min-max heap $\id{mmheap}$ is of size $m$.}
    \label{alg_explore}
    \begin{algorithmic}[1] 
        \Procedure{Explore}{$p, \id{mmheap}, r, \mE, m$}
            \State $\id{minheap} \gets \{\}$       \Comment{stores triplets, sorted by first value}
            \For {$idx \in \{\Call{FreeIdxs}{p}\}$} \Comment{Free indices of $p$}
                \State $\id{cheap} \gets \mE[idx][1]$ \Comment{Cheapest substitution}
                \State $\id{newtuple} \gets (\id{cheap} + \Call{Dist}{p,r}, \id{idx}, 1) $
                \State $\Call{Push}{\id{newtuple}, \id{minheap}}$
            \EndFor
                    \State $(\id{msb}, \id{fid}, \id{sid}) \gets \Call{FindMin}{\id{minheap}}$ 	
                    \State \Comment{\id{msb} is the minimum substitution cost, \id{fid} the free index, \id{sid} the substitution index}
            \If {!$\Call{IsFull}{\id{mmheap}}$} \Comment{$\lvert \id{mmheap}\rvert < m$}
                \Repeat
                    \State $\Call{MakeNewSubK}{p, \id{minheap}, \id{mmheap}, \mE}$
                     \State $(\id{msb}, \id{fid}, \id{sid}) \gets \Call{FindMin}{\id{minheap}}$
                \Until{$\Call{IsFull}{\id{mmheap}}$}
            \Else
                 \State $\id{max} \gets \Call{FindMax}{\id{mmheap}}$
                \While{$\id{msb} < \Call{Dist}{\id{max}, r}$} 
                        \State $\Call{MakeNewSubK}{p, \id{minheap}, \id{mmheap}, \mE}$
                        \State $(\id{msb}, \id{fid}, \id{sid}) \gets \Call{FindMin}{\id{minheap}}$
                		\State $\id{max} \gets \Call{FindMax}{\id{mmheap}}$
                 \EndWhile
            \EndIf
        \EndProcedure
    \end{algorithmic}
\end{algorithm}

\subsection{Overlapping with Substitute K-mers}
\label{subsubsec:methods:find_overlaps_withsubstitutes}

The matrix formulation we used in Section~\ref{subsec:methods:find_overlaps} can be elegantly extended to find overlaps with substitute \kmer{s}.
The naive approach is to modify matrix $\mA$ with all the substitute \kmer{s}.
However, this would be costly.
A protein of length $L$ has $L-k+1$ $k$-mers in it.
Given that $L$ is often around $100-1000$, this itself is not a problem.
However, modifying $\mA$ to include all substitute $k$-mer would increase the number of nonzeros in each row to potentially $m\times(L-k+1)$, severely increasing the computational cost and memory consumption.

Instead, we use a second matrix that encodes the \kmer{}-to-substitute-\kmer{} mappings.
This matrix $\mS$ is at most of dimensions $\lvert \Sigma \rvert^k \times \lvert \Sigma \rvert^k$ but its sparsity is controlled at $m$ nonzeros per row.
The matrix $\mS$ does not need to be binary, and we can encode substitution costs in it.
With this modification, our overlapping computation becomes $\mA \mS \mA\transpose$.
Importantly, this requires a new semiring for $\mA \mS$ and a slight change to the existing semiring for $(\mA \mS) \mA \transpose$.
This new semiring chooses the closest \kmer{} for a substitute \kmer{} when there is more than one \kmer{} in a given read that has the same substitute \kmer{}.
For example, suppose a given sequence $i$ has \kmer{s} $k_p$ and $k_q$ at
$\mA_{ip}$ and $\mA_{iq}$.
Also, assume a substitute \kmer{} $k_s$ is common to both $k_p$ and $k_q$
in $\mS$ but each with distances $d_{ps}$ and $d_{qs}$.
Even though substitute \kmer{s} do not actually appear in a given sequence, we record the starting position of their closest \kmer{} as their location.
That is, if $d_{ps} \le d_{qs}$ we would store the position of $k_p$ as the
starting position of $k_s$ and vice versa.
We omit the technical details for brevity.

\subsection{Sparse Matrix Storage}
\label{subsec:sparse-format}
\ac{CombBLAS} supports the doubly compressed sparse column (DCSC)
format~\cite{Buluc2008} (among others) for storing sparse matrices locally on each process.
DCSC is designed for the representation of hypersparse matrices, in which
the number of nonzeros is smaller than the number of rows or columns.
It is an efficient format for hypersparse matrices in terms of space as it
avoids storing pointers of empty columns.
%

\begin{algorithm}[t]
  \small
    \caption{Create and insert a new substitute \kmer{} to an existing set of
    nearest \kmer{s}, only modifying the free index $\id{fid}$.}
    \label{alg_new_subk}
    \begin{algorithmic}[1] 
        \Procedure{MakeNewSubK}{$p, \id{minheap}, \id{mmheap}, r, \mE$}
            \State $(\id{msb}, \id{fid}, \id{sid}) \gets \Call{ExtractMin}{\id{minheap}}$ 
            \State $\id{subk} \gets p$ \Comment{initialize substitute $k$-mer}
            \State $\id{bid} \gets \Call{IndexOf}{\id{subk}[\id{fid}]}$ \Comment{Base to index}
            \State $\id{subk}[\id{fid}] =  \mE[\id{bid}][\id{sid}]\mathtt{.base}$ \Comment{Replace that base}
            \State $\Call{DeleteFreeIdx}{\id{fid}, \id{subk}}$
            \If{$\Call{IsFull}{\id{mmheap}}$}
                \State $\Call{ExtractMax}{\id{mmheap}}$
            \EndIf
            \State $\Call{Push}{\id{subk}, \id{mmheap}}$
            \State $\id{sid} \gets \id{sid}+1$ \Comment{Get next cheapest substitution}
            \State $\id{newtuple} \gets (\mE[\id{fid}][\id{sid}]+\Call{Dist}{p,r}, \id{fid}, \id{sid})$
            \State $\Call{Push}{\id{newtuple}, \id{minheap}}$ %
        \EndProcedure
    \end{algorithmic}
\end{algorithm}

The size of the \kmer{} space in \ac{\swname} is $\lvert \Sigma \rvert^k$, where
$\lvert \Sigma \rvert = 24$.
Even for a relatively small $k$, this results in a large column count
in $\mA$ and row/column counts in $\mS$.
Moreover, as these matrices are distributed by CombBLAS among processes
in a 2D manner (Section~\ref{subsec:impl:data_partitioning}), the columns of submatrices
stored in each process become increasingly empty
as the process count increases.
For example, for the \texttt{Metaclust50-1M} dataset we use in our experiments
(which contains 1 million sequences) and for a \kmer{} size of 6, $\mA$
is a 1M
$\times$ 244M matrix containing 108M nonzeros, and $\mS$ is a 244M $\times$ 244M
matrix containing 611M nonzeros.
These matrices respectively contain 0.44 and 2.50 nonzeros per column, and when
they are distributed among multiple processes, the number of nonzeros per column gets even
smaller in the submatrices stored.
Hence, to scale to thousands of processes, \ac{\swname} stores submatrices in the DCSC format.

\subsection{Alignment of Overlapping Sequences} \label{subsec:methods:alignment}

\ac{\swname} supports two alignment modes for sequence pairs detected in the first step: \ac{XD}~\cite{Altschul1997} and \ac{SW} alignment~\cite{smith1981identification}.

In \ac{XD}, \ac{\swname} initiates the alignment starting from the position of the shared \kmer{s} and extending it in both directions until the end of the sequences using gapped x-drop.
Since we store up to two shared \kmer{s}, the alignment is performed starting from of both of them separately.
The alignment with the best score and that passes the similarity thresholds is retained.

The \ac{SW} alignment computes a local alignment starting from the beginning of each sequence. 
Even though the alignment computations initiates at the beginning of the sequences and it is performed until the end, only the local alignment with higher score between the two sequences is returned.
In \ac{SW}, scores cannot assume negative values.
In this case, our seed position is ignored and the seed is merely used to mark the two sequences as potentially related and worth aligning.
One advantage of this method over seed-and-extend is that alignment quality does not depend on the seeds we found.
The implementation of \ac{XD} and \ac{SW} is offloaded to the \ac{SeqAn} C++ library~\cite{pmid:18184432}.




\subsection{Sequence Similarity Filter}
\label{subsec:methods:similarity_filter}

The last stage of the pipeline is the alignment post-processing, which
includes filtering out sequence pairs that fall below certain quality metrics
provided to \ac{\swname}.
From our experience, these metrics typically include vetoing pairs with alignment similarity less than 30\% and length coverage less than 70\%.

\begin{figure}[t]
  \centering
  \vspace{-1.0em}
    \includegraphics[width=0.65\columnwidth]{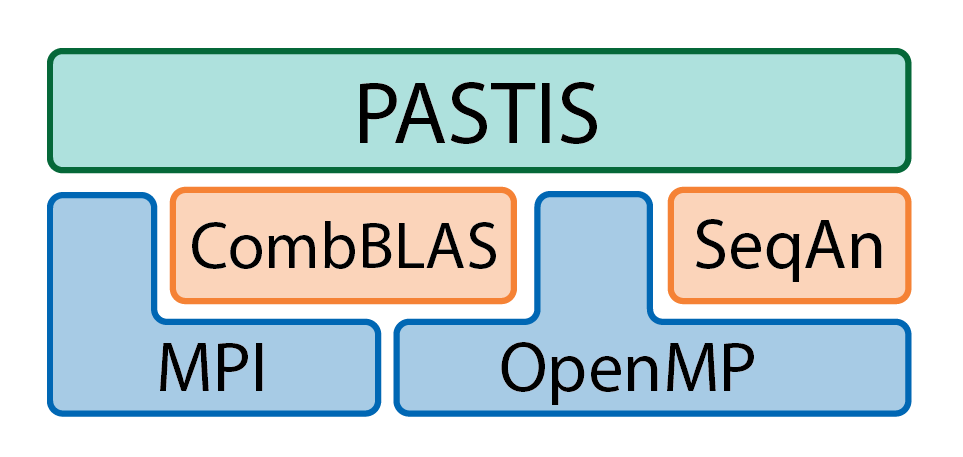}
    \vspace{-1.0em}
    \caption{PASTIS software stack. PASTIS implicitly uses MPI and/or OpenMP
      through CombBLAS and SeqAn libraries. It also explicitly uses both
      parallel programming paradigms in certain stages.}
    \label{fig:pastis-stack}
\end{figure}

\section{\ac{\swname} Distributed Implementation} \label{sec:impl}

\ac{\swname} provides a single scalable distributed implementation of the
pipeline illustrated in Figure~\ref{fig:homology_detection_pipeline}.
The connections found in the \ac{PSG} are oblivious to the number of processes
used to parallelize \ac{\swname}.
This is an important aspect of our tool as it allows reproducible results under
different configurations.
Some of the suffix array based existing tools such as LAST, in contrast, may
produce configuration specific results depending on the physical resource
limitations.

The current implementation of \ac{\swname} uses \ac{MPI} as its underlying
parallel communication library.
We assume \ac{\swname} is run with $p=q^2$ number of parallel processes, where
$q$ is a positive integer.
The requirement to have a square number of processes is due to 2D domain
decomposition that happens during sparse matrix creation and multiplication
using the \ac{CombBLAS} library.
The following sections describe the implementation details along with some of
the computation and communication optimizations.
Figure~\ref{fig:pastis-stack} shows the libraries and parallel programming
paradigms \ac{\swname} relies on.

Among the two libraries utilized by \ac{\swname}, \ac{CombBLAS} supports hybrid
MPI/OpenMP parallelism while \ac{SeqAn} supports shared-memory parallelism with
OpenMP.
By utilizing these libraries, \ac{\swname} implicitly makes use of the inherent
parallelism in them.
Apart from those, \ac{\swname} explicitly makes use of MPI/OpenMP hybrid
parallelism as well.
For an example, in the preparation of batches of pairwise alignments for
\ac{SeqAn}, it uses OpenMP threads.
Then, when \ac{SeqAn} completes these alignments, the threads at each process
again process the output information in parallel to gather necessary statistics
for forming the similarity graph.
In this way, OpenMP is used both implicitly and explicitly in \ac{\swname}.

\subsection{Data Partitioning} \label{subsec:impl:data_partitioning}

The input to \ac{\swname} is a set of protein sequences in \ac{FASTA} format.
We use parallel file I/O to read independent chunks of this file.
Each process gets the size of the file and then divides this size evenly by the
number of processes.
This gives a begin and end location, a chunk, to be read by each process in
parallel.
It is common that a file chunk read with such splitting may not start and end at
sequence margins, so we read a user defined extra amount of bytes in each
process.
With this approach, each process will ignore any partial sequences at the start
of its chunk and read until the end of the last sequence in chunk with possibly
reading over on the extra bytes at the end.

\begin{figure}[t]
    \centering
    \includegraphics[width=0.55\columnwidth]{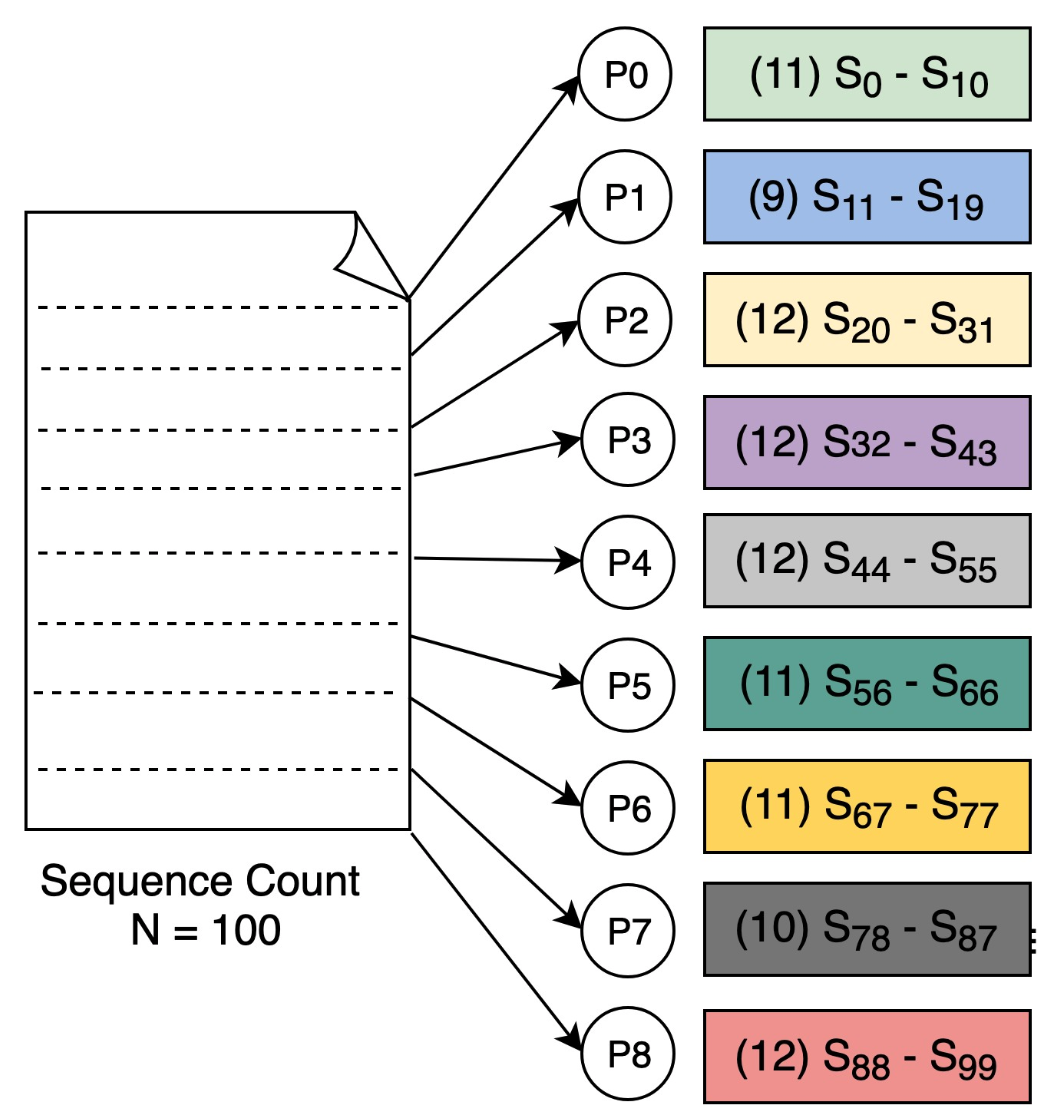}
    \vspace{-1em}
    \caption{Input sequence partitioning in \ac{\swname}}
    \label{fig:data_partition}
\end{figure}

Figure~\ref{fig:data_partition} shows an example of how \ac{\swname} would
partition $\mathrm{100}$ sequences among $\mathrm{9}$ processes.
Note how some processes have different number of sequences due to the length variation among
proteins. 
This variation does not cause any load imbalance because the runtime of the parallel I/O section depends on the total length of the sequences that is being read and parsed,
which is exactly what our implementation balances by choosing process boundaries via assigning each process an equal number of bytes as opposed to an equal number of sequences.
%
%
%
The processes then start communicating sequences to fit into a 2D distribution
(Section~\ref{subsec:impl:overlapping_comm}). Each process is responsible for a predefined range of protein IDs in this 2D grid, ensuring
load balance by construction.
%

Internally, \ac{\swname} stores a pointer to the character buffer of its
sequences in each process for efficient storage without converting to a separate
data structure.
It records sequence identifier and data start offsets, so any sequence can be
accessed using a local or global index.
A parallel prefix sum of sequence counts are computed cooperatively by all
processes, so that each process is aware what sequences are stored by which
processes.

\subsection{Seed Discovery} \label{subsec:impl:seed_finding}

As explained in Section~\ref{subsec:methods:find_overlaps}, seed discovery requires the creation and multiplication of $\mA$ with its transpose.
The columns $\mA$ present a direct mapping to \kmer{s}.
Therefore, we compute a unique number for each \kmer{} as follows.

We index each base in the protein alphabet uniquely from $\mathrm{0}$ to $\mathrm{23}$.
Then each base gets a number as $b24^{i}$, where $b$ is the index of the base in the alphabet and $i$ is the zero-based position of the base in the \kmer{} from right to left.
For example, under the $\mathtt{ARNDCQEGHILKMFPSTWYVBZX*}$ alphabet, the $\mathrm{3}$-mer $\mathtt{RCQ}$ will be assigned the id $\mathrm{1\cdot24^2+4\cdot24^1+5\cdot24^0 = 677}$.
Importantly, we only perform this for \kmer{s} present in sequences and not the entire \kmer{} space.


\begin{figure}[t]
    \centering
    \includegraphics[width=0.65\columnwidth]{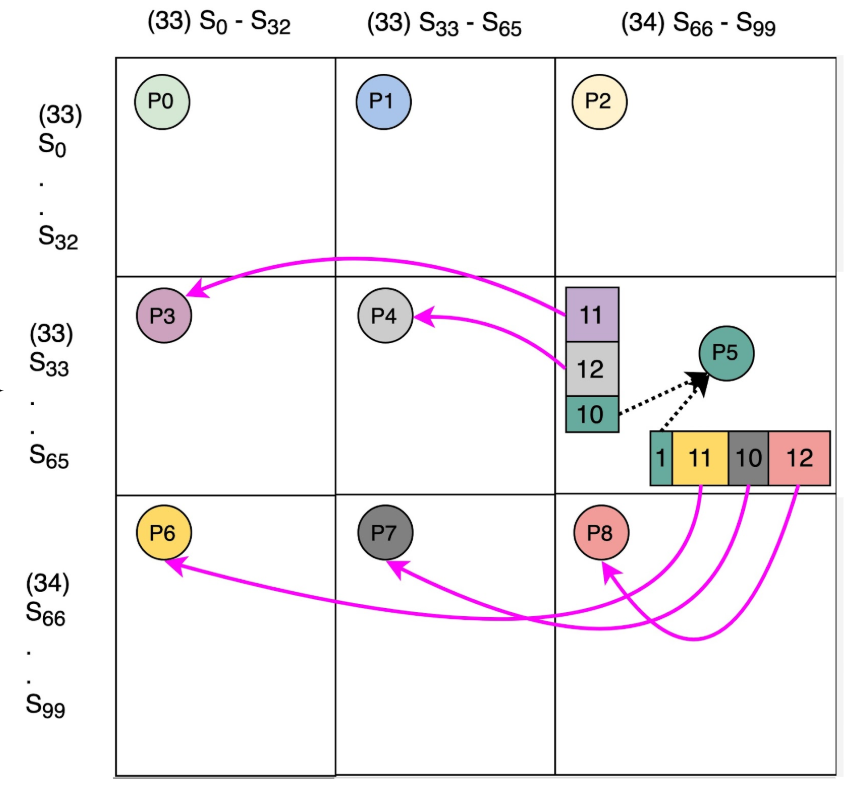}
    \vspace{-0.75em}
    \caption{Remote sequence requests from $\mathrm{P5}$ to other processes.}
    \label{fig:nbr_data}
\end{figure}

\subsection{Overlapping Communication} \label{subsec:impl:overlapping_comm}

\ac{CombBLAS} uses a 2D decomposition of matrices, so both $\mA$ and $\mB$ will
be distributed onto a process grid of $\sqrt{p} \times \sqrt{p}$, where $p$ is
the total number of processes.
This decomposition is especially important for $\mB$ as nonzero elements of this
matrix indicate sequence pairs that need to be aligned but a particular process
may not have the corresponding sequences in its partition.

To clarify, consider the example in Figure~\ref{fig:nbr_data}.
Here, we show the decomposition of matrix $\mB$ over a $\mathrm{3\times3}$ process grid identified as $\mathrm{P0}$ through $\mathrm{P8}$.
If we look at $\mathrm{P5}$, for example, it needs sequences, $\mathrm{S_{33}}$ through $\mathrm{S_{65}}$ (row sequences), and $\mathrm{S_{66}}$ through $\mathrm{S_{99}}$ (column sequences).
While, in practice, it may not need all of these sequences as not each pair within these ranges will have shared \kmer{s}, these represent the entire space of sequences $\mathrm{P5}$ would need in the worst case.
However, based on the linear decomposition of sequences, $\mathrm{P5}$ only has sequences $\mathrm{S_{56}}$ through $\mathrm{S_{66}}$ available locally.
Any remaining sequences it would need has to be fetched from other
processes that contain them.

In fetching remote sequences, a process has the option to either wait until $\mB$ is computed to figure out the sequences it would need or request the full range of sequences it might need.
We observe that with a parallelism of $p$, each process has to store $2n/\sqrt{p}$ sequences, at the most.
Given the memory available in today's machines and with a reasonable $p$, we note it is feasible to store this many sequences locally.
The advantage of this decision is that it allows to perform remote sequence fetching in the background while the operations such as seed discovery, matrix creation, and matrix multiplication are being performed.

Figure~\ref{fig:overlapping_comm} illustrates the implementation of overlapping communication in \ac{\swname}.
Immediately after reading local sequences, each process computes the ranks that it has to request from as well as the ranks that will request from it.
It will then proceed to issue the necessary $\mathtt{MPI\_Irecv}$ and $\mathtt{MPI\_Isend}$ calls to initiate the remote sequence exchange.
After computing $\mB$, an $\mathtt{MPI\_Waitall}$ guarantees that each process has received the sequences it needs.

\subsection{Moving Computation to Data} \label{subsec:impl:com_to_data}

Once $\mB$ is computed, the nonzero sequence pairs in its upper triangular portion needs to be aligned.
If this were done na\"ively, a $\sqrt{p} (\sqrt{p}-1)/2$ number of processes would sit idle.
Alternatively, one could redistribute the pairs in the upper triangular portion among all processes but that would incur additional communication cost.

\begin{figure} [t]
    \centering
    \includegraphics[width=0.85\columnwidth]{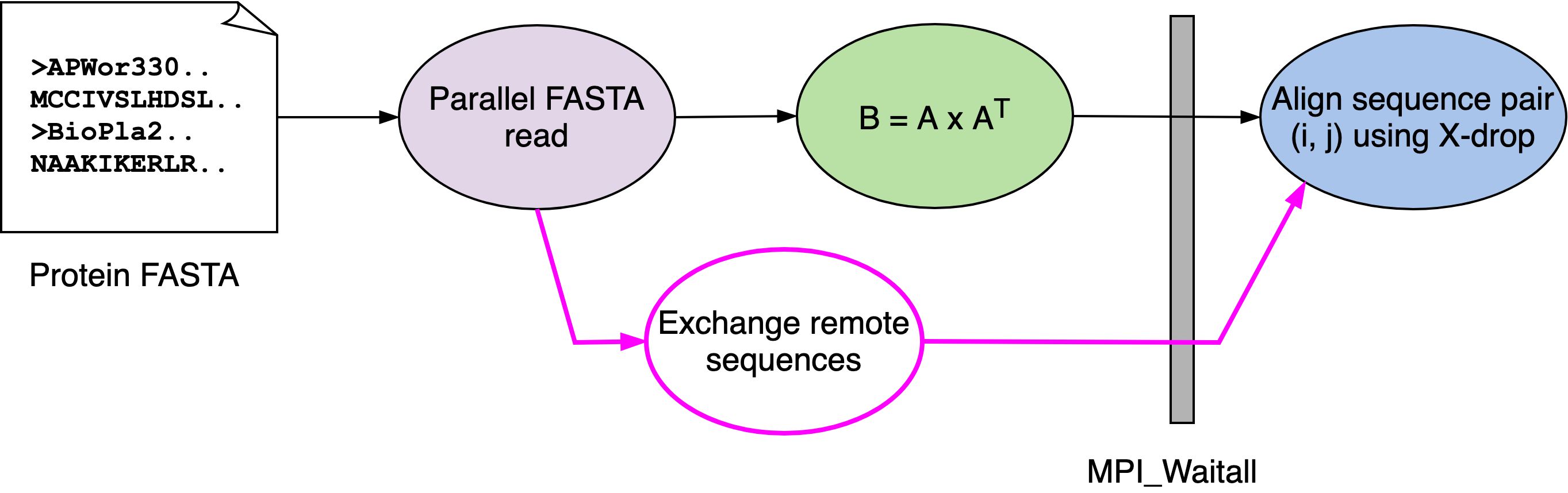}
    \vspace{-1em}
    \caption{Overlapping communication and computation in \ac{\swname}.}
    \label{fig:overlapping_comm}
\end{figure}

We can avoid both of these and move computation to data for free since $\mB$ is symmetric.
This is shown in Figure~\ref{fig:com_distribution}.
Note all processes except the off-diagonal ones in the last row and column of
the figure are square blocks.
Despite their dimensions, we see that all the pairs in the upper triangular portion of
$\mB$ are equivalent to the collection of pairs in individual upper triangular portions of each
process block.

For example, the matrix at the left of Figure~\ref{fig:com_distribution} shows that
$\mathrm{P5}$ should compute all nonzero pairs in its row and column
sequence ranges.
At right, we show that this is equivalent to $\mathrm{P5}$ computing
nonzero pairs in its upper triangular and $\mathrm{P7}$ computing nonzero pairs
in its upper triangular.
Also note that the diagonal entries of each block will only be computed by
processes on or above the main diagonal, i.e., $\mathrm{P0, P1, P2, P4, P5, P7}$.

\section{Evaluation}
\label{sec:experiments}

We present our evaluation in two categories: parallel performance and
relevance.
We use different set(s) of data in each.
In the former category, we use subsets of
\texttt{Metaclust50}~\cite{pmid:19341454} dataset.
Originally, this dataset has a total of 313 million sequences and we use random
subsets of different sizes in our evaluation: 0.5, 1, 1.25, 2.5, and 5 million
sequences, depending on the setting of the experiment.
We indicate the subset size within the evaluated dataset, e.g.,
\texttt{Metaclust50-0.5M} for a subset of 0.5 million sequences.
%
%
In the latter category, we use the curated (a combination of automatic and
manual curation) dataset \texttt{SCOPe} (Structural Classification of Proteins
-- extended)~\cite{pmid24304899} to measure precision and sensitivity.
From the original set of 243,813 proteins in this dataset, we pick 77,040 unique proteins.
\texttt{SCOPe} dataset has 4,899 protein families.
%
In the relevance evaluation, we compare \ac{\swname} against
\ac{MMseqs2}~\cite{steinegger2017mmseqs2}, a many-against-many sequence
searching and clustering software for large protein and nucleotide sequence
sets, and LAST~\cite{pmid:21209072}, an aligner with an emphasis on discovering
weak similarities.
The evaluation and discussions for the two categories are presented in
Section~\ref{sec:perf}~and~\ref{sec:prec-recall}, respectively.

We conduct our evaluations on NERSC Cori system - a Cray XC40 machine.
Each Haswell node on this system consists of two 2.3 GHz 16-core Intel Xeon
E5-2698 v3 processors and has a 128 GB of total memory.
Each KNL node consists of a single 1.4 GHz 68-core Intel Xeon Phi 7250 processor and has 96 GB of total memory.
We use Haswell nodes in our comparisons against \ac{MMseqs2} and LAST, while we
use KNL nodes in analyzing scalability of \ac{\swname}.
The KNL partition of Cori has more nodes and hence it allows larger-scale
experiments.
Given \ac{MMseqs2} does not support AVX512, the comparison against it is run on Haswell nodes with AVX2 instructions for a fair comparison.

Both \ac{\swname} and \ac{MMseqs2} support hybrid MPI and OpenMP parallelism.
For both tools, we assign each node a single MPI task with as many threads as there are cores on the node.
%
%
Both tools are compiled with \texttt{gcc~8.3.0} and the \texttt{O3} flag.
In our evaluations, both \ac{MMseqs2} and the \ac{SeqAn} library used for
alignment in \ac{\swname} make use of vectorization with \ac{AVX2}, while LAST
uses SSE4.
%

\begin{figure} [t]
    \centering
    \includegraphics[width=0.80\columnwidth]{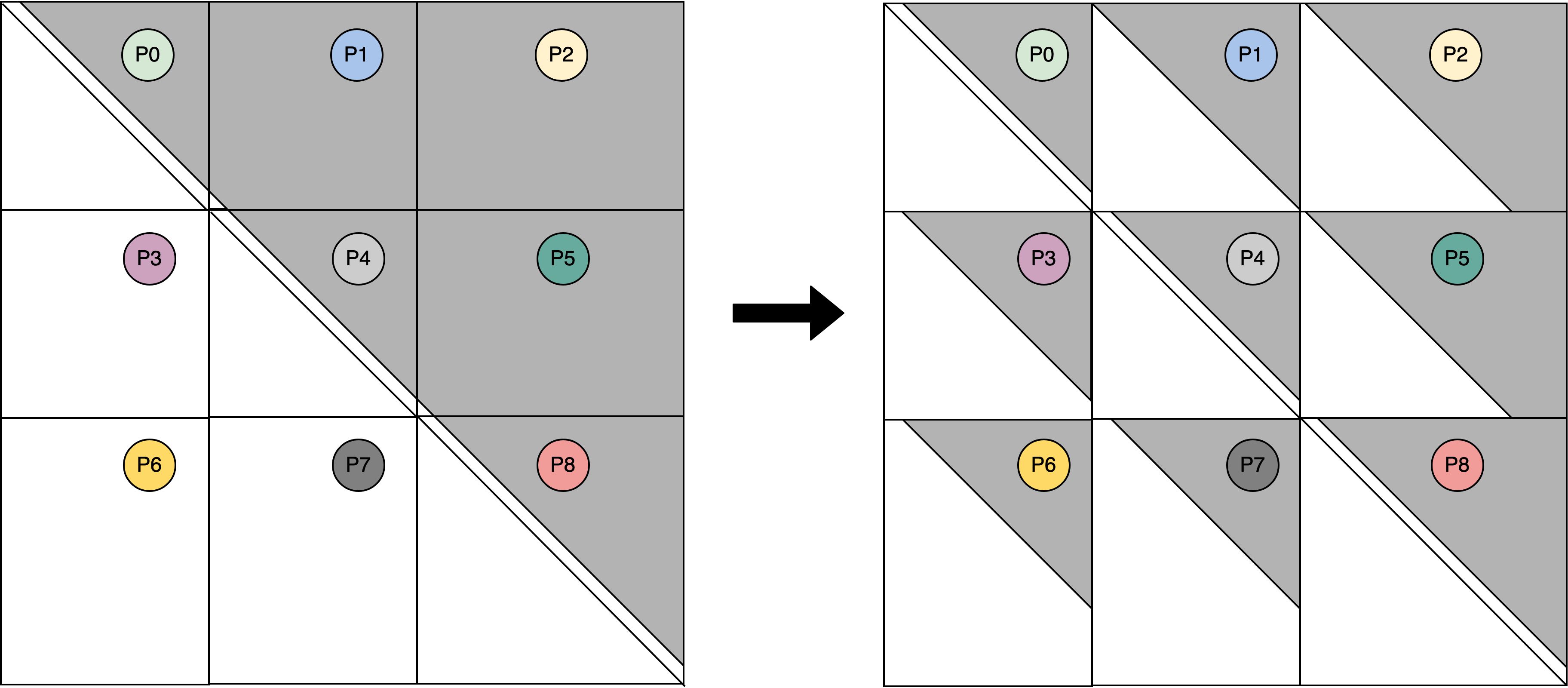}
    \caption{Distribution of computation over data.}
    \label{fig:com_distribution}
\end{figure}

Both alignment modes described in Section~\ref{subsec:methods:alignment} are
used in our evaluation of \ac{\swname}.
%
To indicate these modes, we add the alignment modes' abbreviations as a suffix
to our tool, e.g., \ac{\swname}-\ac{SW} or \ac{\swname}-\ac{XD}.
For \ac{XD} alignment, we use an x-drop value of 49.
We use a \kmer{} size of 6 for \ac{\swname}.
%
For \ac{\swname}, we also evaluate a parameter called common \kmer{} threshold.
When this parameter is set to t, we eliminate (i.e., do not perform alignment on) read pairs that share t or fewer $k$-mers.
For exact \kmer{s} we use a common \kmer{} threshold of 1 and for substitute
\kmer{s} we use a common \kmer{} threshold of 3.
The variants of \ac{\swname} with this parameter are indicated with CK suffix.
%
During the pairwise alignment, we use the BLOSUM62 substitution
matrix~\cite{pmid:1438297} with a gap opening cost of 11 and a gap extension
cost of 1.
For \ac{MMseqs2} and LAST, we use the default settings except for the
sensitivity, identity score and coverage thresholds, which change according to
the experiment setup as described later.
Note that LAST's parallelism is constrained to a single node.
Nonetheless, we include it in our evaluation mainly for sensitivity comparison,
although we also report its single-node runtime performance.
%

\begin{figure}[t]
  \centering
    \includegraphics[width=0.565\columnwidth]{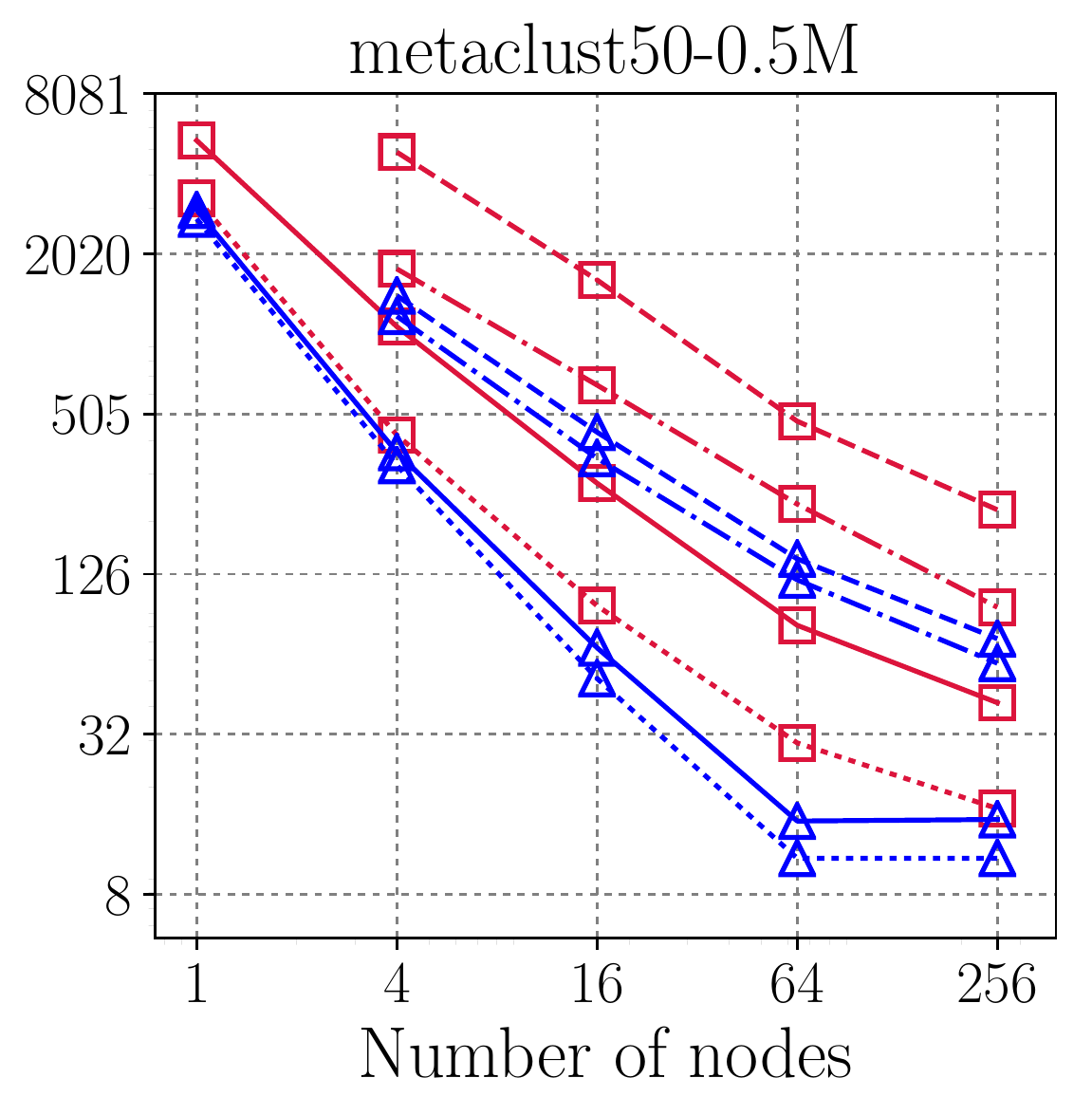}\hspace{-0.71em}
    \includegraphics[width=0.445\columnwidth]{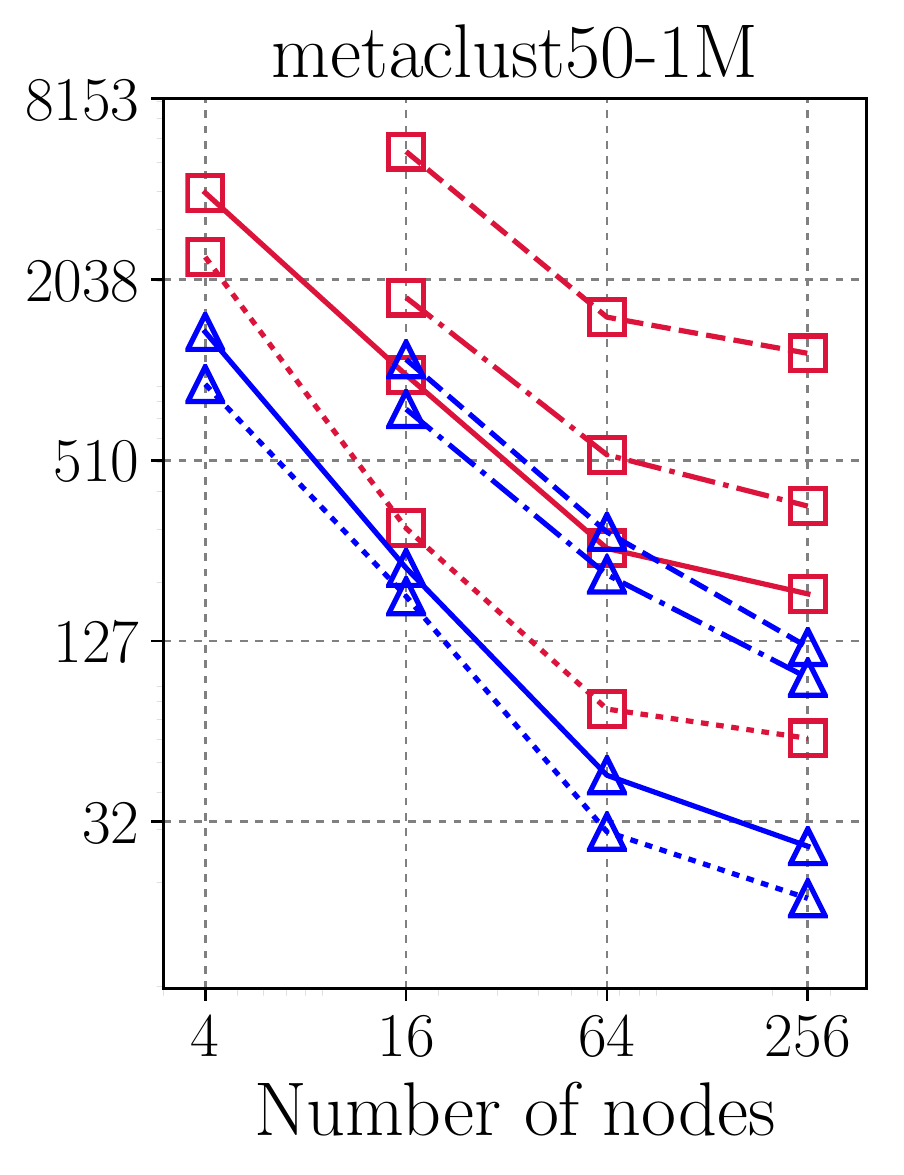}
    \vspace{-0.2em}
    \includegraphics[width=0.75\columnwidth]{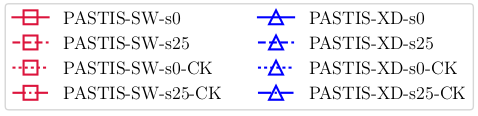}
    \caption{Runtime of \ac{\swname} variants on two datasets. The y-axis
      denotes the execution time in seconds.}
    \label{fig:pastis-variants}
\end{figure}

\subsection{Parallel Performance}
\label{sec:perf}

First, we evaluate parallel performance of \ac{\swname} by comparing it against
another distributed memory aligner \ac{MMseqs2} and shared memory aligner LAST.
Then, we focus on scaling aspects and investigate how different components of
\ac{\swname} scale.

\noindent
{\bf Comparison against \ac{MMseqs2} and LAST.}
For the comparison we use two datasets with 0.5 and 1 million sequences.
%
%
We vary the number of nodes from 1 to 256, increasing it by a factor of 4.
Additionally, we evaluate a number of different options for \ac{\swname}.
For the alignment, we test \ac{SW} and \ac{XD}.
For substitute \kmer{s} we use the values of 0 (i.e., there is no $\mS$ matrix)
and 25.
These are indicated with suffixes to \ac{\swname}.
For \ac{MMseqs2}, we test out three different sensitivity settings (parameter
$s$): low (1.0), default (5.7), and high (7.5).
A smaller sensitivity value should result in a faster execution for
\ac{MMseqs2}.
For LAST, we use 100 for the maximum initial matches per query position.
This parameter controls the sensitivity and the higher it is, the more time LAST
takes.
Figure~\ref{fig:pastis-variants} compares \ac{\swname} variants among themselves
and Figure~\ref{fig:vs-mmseqs} shows the execution time of the fastest variant
of \ac{\swname}, \ac{MMseqs2}, and LAST.
The missing data points for \ac{\swname} are due to running out of memory.

\begin{figure}[t]
  \centering
    \includegraphics[width=0.515\columnwidth]{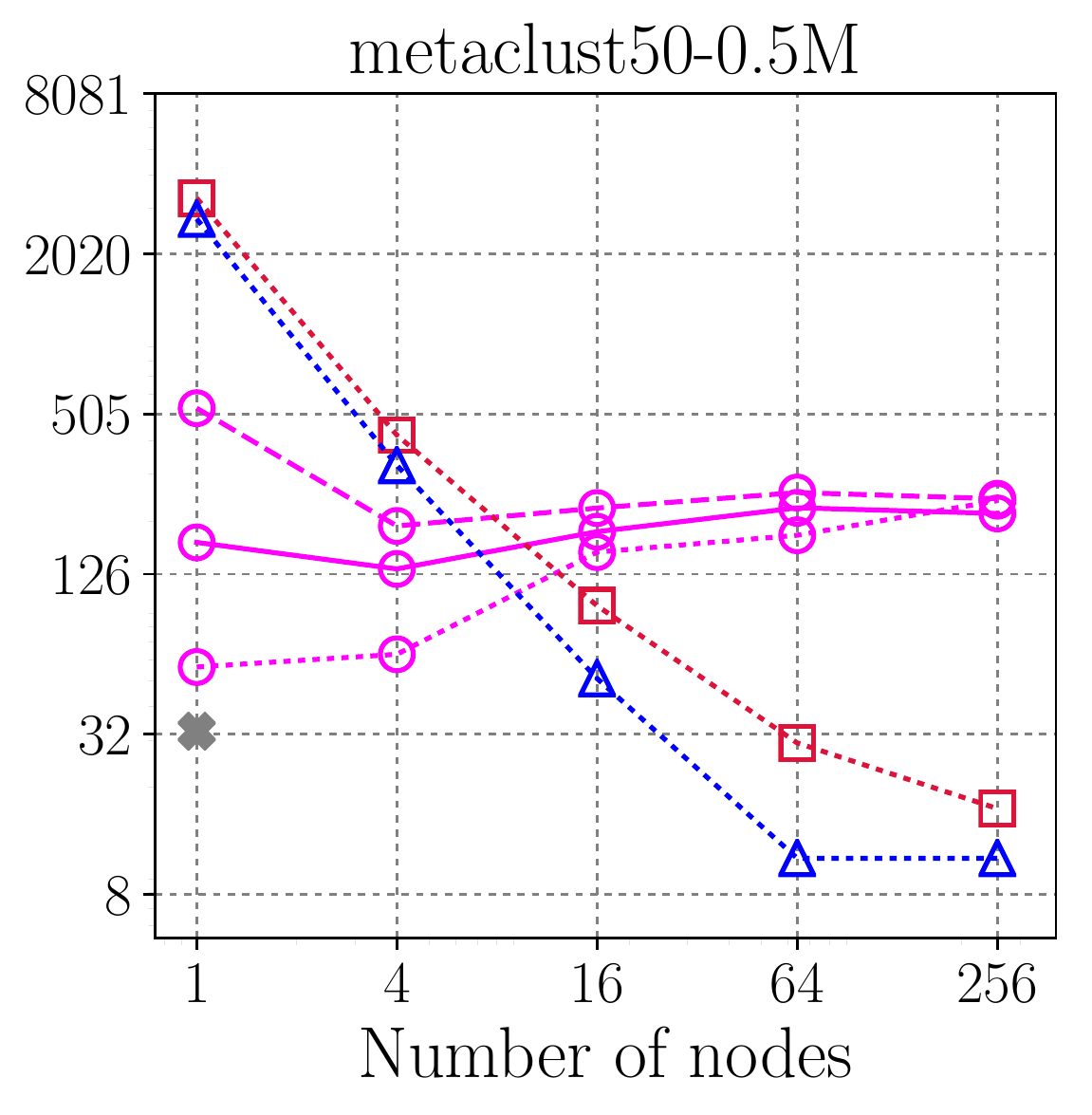}\hspace{-0.71em}
    \includegraphics[width=0.495\columnwidth]{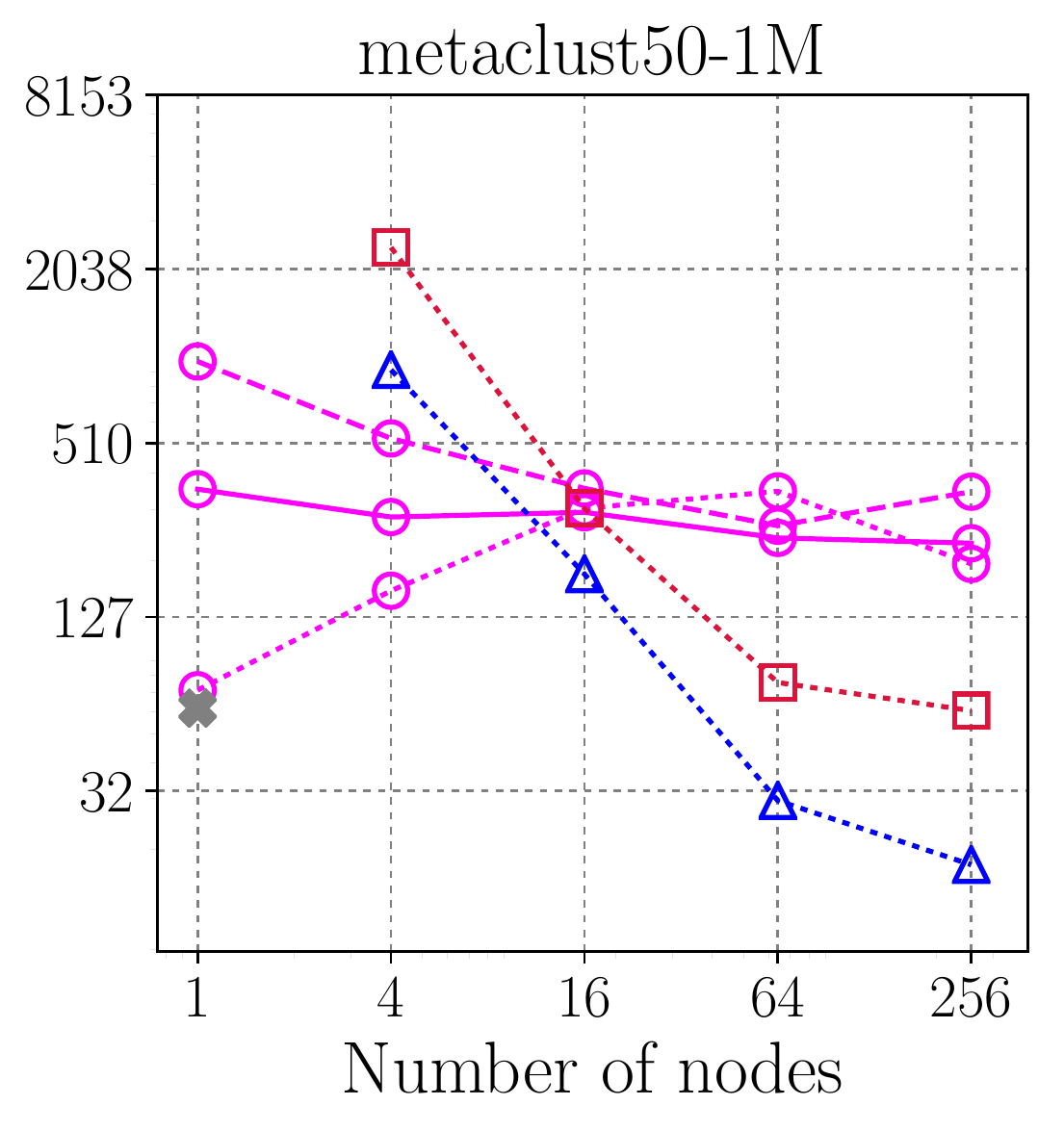}
    \vspace{-0.2em}
    \includegraphics[width=0.75\columnwidth]{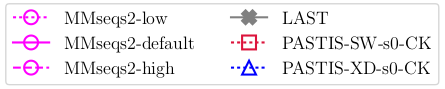}
    \caption{Runtime of \ac{\swname} vs. \ac{MMseqs2} on two datasets. The
      y-axis denotes the execution time in seconds.}
    \label{fig:vs-mmseqs}
\end{figure}

When we compare different variants of \ac{\swname} in
Figure~\ref{fig:pastis-variants}, we can see that using substitute \kmer{s}
increases the execution time as expected because we add the substitution matrix
to the sparse matrix computations and increase the number of alignments.
%
For instance, in \texttt{Metaclust50-0.5M}, the number of alignments performed
with exact \kmer{s} is 399 million whereas with 25 substitute \kmer{s} it is 3.5
billion -- amounting to a factor of 8.7$\times$ in the number of alignments.
\ac{XD} is substantially faster than \ac{SW}, without any significant change in
accuracy (we discuss this in Section~\ref{sec:prec-recall}).
The \ac{\swname} variants that use the common \kmer{} threshold are faster as
they perform fewer alignments than their counterparts.

Figure~\ref{fig:vs-mmseqs} shows that \ac{\swname} is slower than \ac{MMseqs2}
for small node counts but, due to its better scalability, it is able to close
the performance gap rather quickly.
\ac{\swname}-\ac{XD}-s0-CK runs faster than \ac{MMseqs2} in both datasets
starting around 16 nodes.
\ac{\swname} often scales favorably.
%
%
%
We investigated the unscalable behavior of \ac{MMseqs2} and found out that
although the computations scale well, the processing after running the
alignments constitutes bulk of the time and causes a bottleneck.
In that processing stage, \ac{MMseqs2} probably gathers alignment results from
other nodes in order to write the output using a single process, which is
handled in parallel in \ac{\swname}.
Among different variants of \ac{MMseqs2}, \ac{MMseqs2}-low runs faster as
expected in a single-node setting.
However, \ac{MMseqs2}-high scales somewhat better as it is more compute-bound
than the two other variants.
LAST's single-node performance is better than three variants of \ac{MMseqs2}.

\begin{figure}[t]
    \centering
    \includegraphics[width=0.50\columnwidth]{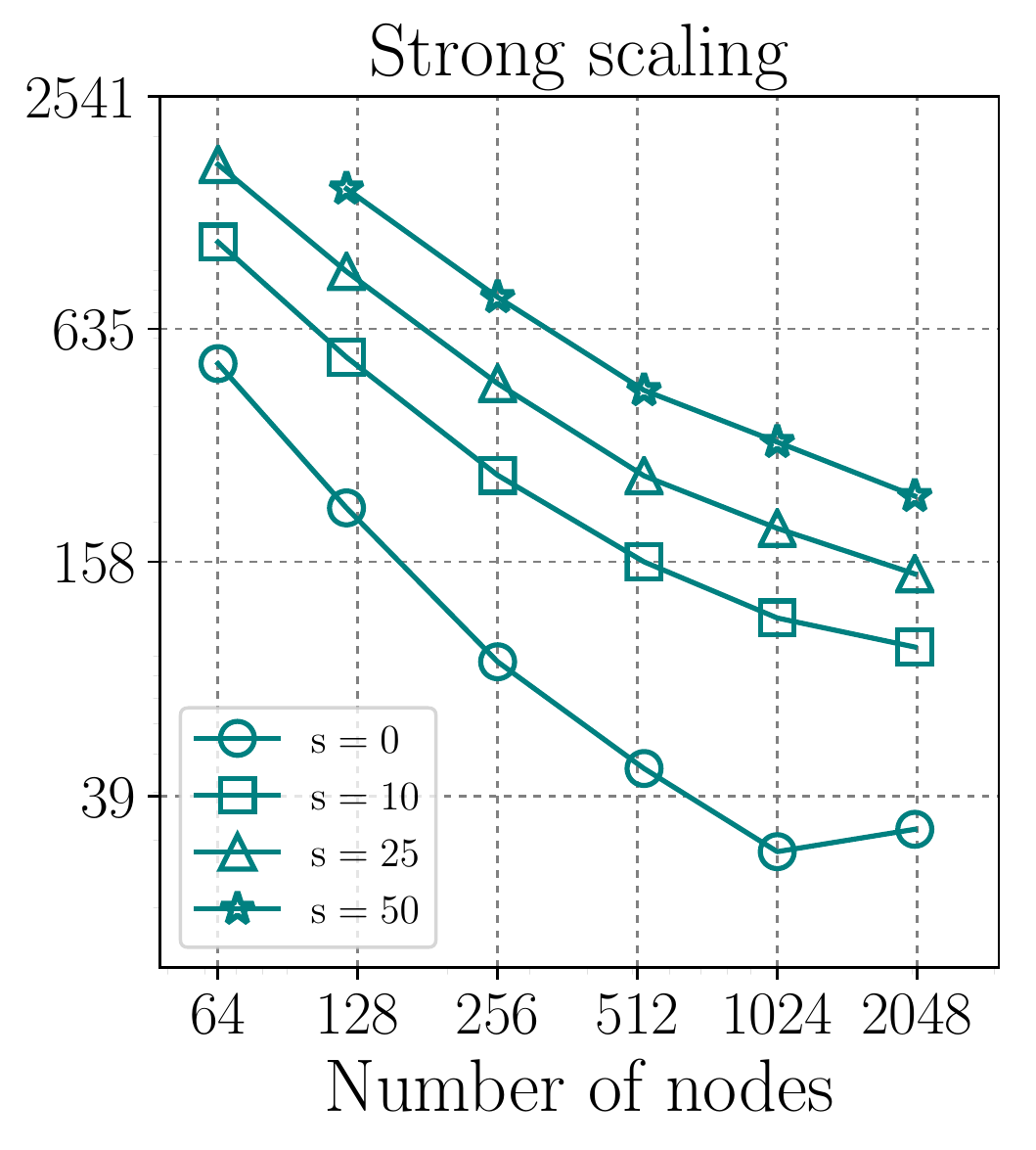}\hspace{-0.71em}
    \includegraphics[width=0.50\columnwidth]{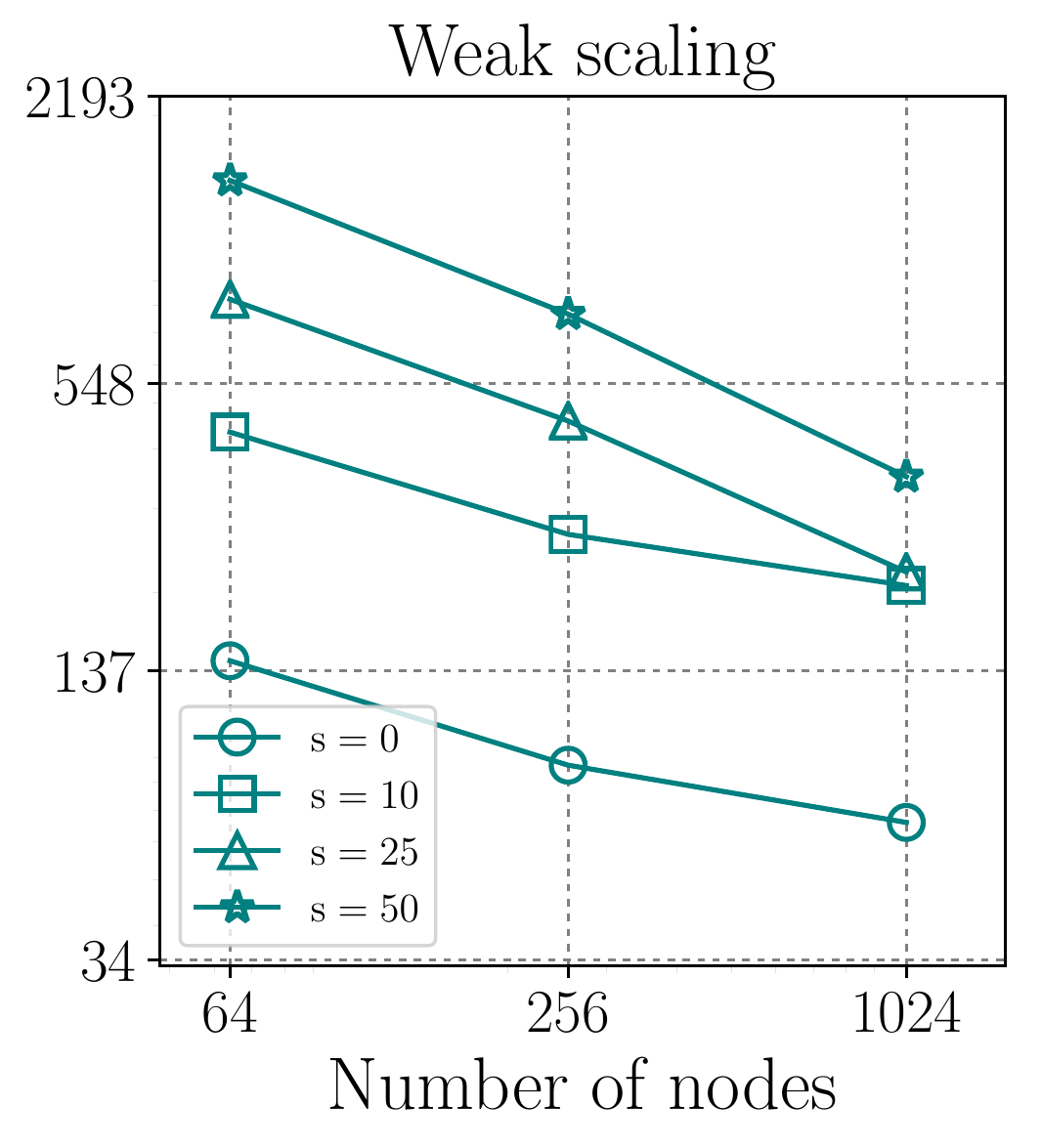}
    \caption{Strong-scaling (left) and weak-scaling (right) performance of
      \ac{\swname}. The y-axis is the execution time in seconds.}
    \label{fig:strong-scaling}
\end{figure}

The percentage of time spent in aligning pairs in \ac{\swname} is presented in
Table~\ref{tb:algn-perc}.
The alignment percentages are higher for \ac{SW} as it is more expensive than
\ac{XD}.
Increasing the number of sequences from 0.5 million to 1 million causes a
quadratic increase in the number of aligned pairs (will be discussed shortly),
while sparse matrix operations usually scale linearly.
Hence, the percentage of time spent in alignment tends to increase with
increased number of sequences.

\noindent
{\bf Strong and Weak Scaling.}
%
In this section, we solely focus on the sparse matrix operations and exclude
alignment.
The alignment computations are independent of each other and each process
computes its portion of alignments without requiring any coordination or
synchronization with other processes.
The local alignments at a process are also highly parallel as aligned pairs are
also independent of each other.
Conversely, the sparse matrix computations have different flavors and
accommodate more challenges in terms of parallelization.
Hence, we only focus on the scalability of \ac{\swname} without alignment of
sequence pairs.
The experiments in this section are performed on the KNL partition of Cori to
exploit a larger number of nodes.

On the left side of Figure~\ref{fig:strong-scaling}, we illustrate the strong
scaling behavior on \texttt{Metaclust50-2.5M} with number of substitute \kmer{s}
$\{0, 10, 25, 50\}$, and the number of nodes $\{64, 121, 256, 529, 1024,
2025\}$.
The odd number of nodes is due to perfect square process count requirement in \ac{\swname}.
We choose the perfect square integer closest to the target process count.
Figure~\ref{fig:strong-scaling} shows that using exact \kmer{s} exhibits better
scalability than using substitute \kmer{s} up to 2K nodes.
Substitute \kmer{s} exhibit similar scalability among themselves with an
increase in runtime as the number of substitutions increases.
The substitute \kmer{s} implementation requires formation of the substitution
matrix $\mS$, an additional \ac{SpGEMM}, and the symmetrization of the
similarity matrix.
These components seem to be less scalable compared other components.

On the right side of Figure~\ref{fig:strong-scaling}, we illustrate the weak
scaling behavior.
We use three different datasets: \texttt{Metaclust50-1.25M} at 64 nodes,
\texttt{Metaclust50-2.5M} at 256 nodes, and \texttt{Metaclust50-5M} at 1024
nodes.
We observe that the number of nonzeros in the output matrix increases roughly by
a factor of four when we double the number of sequences.
For example, for 25 substitute \kmer{s}, the output matrices resulting from
using 1.25M, 2.5M, and 5M sequences respectively contain 10.9, 43.3, and 172.3
billion nonzeros.
%
%
The lines in the weak scaling plots have a negative slope, which may seem
unrealistic.
However, not all operations scale with a quadratic factor.
For example, some operations such as communicating sequences and generating
substitute \kmer{s} scale linearly with the number of processes.
Consequently, a weak-scaling line with a negative slope may be expected as we
increase the number of nodes with a factor of four.
On the other hand, a factor of two would result in lines with a positive slope.

\begin{table}[t]
  \centering
  \caption{Alignment time percentage in \ac{\swname}.}
  \scalebox{0.75} {
    \begin{tabular}{l r r r r r r r r r r}
      \toprule
      & \multicolumn{5}{c}{\texttt{Metaclust50-0.5M}} & \multicolumn{5}{c}{\texttt{Metaclust50-1M}} \\
      \cmidrule(r{4pt}){2-6} \cmidrule(r{4pt}){7-11}
      Scheme & 1 & 4 & 16 & 64 & 256 & 1 & 4 & 16 & 64 & 256  \\
      \midrule      
      \ac{\swname}-\ac{SW}-s0  & 49\% & 83\% & 89\% & 91\% & 81\% & - & 73\% & 91\% & 94\% & 71\% \\  
      \ac{\swname}-\ac{SW}-s25 & -    & 80\% & 81\% & 78\% & 75\% & - & -    & 90\% & 89\% & 94\% \\ 
      \ac{\swname}-\ac{XD}-s0  & 7\%  & 54\% & 55\% & 55\% & 52\% & - & 51\% & 59\% & 64\% & 50\% \\
      \ac{\swname}-\ac{XD}-s25 & -    & 31\% & 29\% & 25\% & 27\% & - & -    & 50\% & 40\% & 39\% \\
      \midrule
      \ac{\swname}-\ac{SW}-s0-CK  & 12\% & 60\% & 69\% & 77\% & 64\% & - & 58\% & 71\% & 77\% & 62\% \\  
      \ac{\swname}-\ac{SW}-s25-CK & -    & 44\% & 53\% & 51\% & 48\% & - & -    & 68\% & 66\% & 69\% \\ 
      \ac{\swname}-\ac{XD}-s0-CK  & 1\%  & 48\% & 44\% & 34\% & 33\% & - & 48\% & 47\% & 44\% & 36\% \\
      \ac{\swname}-\ac{XD}-s25-CK & -    & 17\% & 11\% &  6\% &  7\% & - & -    & 25\% & 15\% & 12\% \\
      \bottomrule
    \end{tabular}
  }
  \label{tb:algn-perc}
\end{table}

\noindent
{\bf Dissection Analysis.}
In this section, we examine the time spent in various components of \ac{\swname}
and how these components scale.
As in the previous section, we exclude alignment from our analysis.
We measure the time of 5 different components when exact \kmer{s} are used
and 8 different components when substitute \kmer{s} are used.
Figure~\ref{fig:dissection-all} shows the obtained results.
The components that are specific to substitute \kmer{s} are (i) the formation of
$\mS$, (ii) $\mA \mS$, and (iii) the symmetrization.
The last component is required to make the output matrix symmetric.
The ``fasta.'', ``tr. $\mA$'', and ``wait'' components respectively stand for
reading/processing of fasta data, computing $\mA\transpose$, and waiting for the communication of sequence data to be completed (Section~\ref{subsec:impl:overlapping_comm}).

Figure~\ref{fig:dissection-all} shows that waiting for the sequence transfers
to complete constitutes a considerable portion of the overall time, especially at small node counts.
This component is less pronounced when substitute \kmer{s} are used as other
components take more time while the sequence transfer time stays the same.
For the exact \kmer{s}, the most computationally dominant component is the
\ac{SpGEMM}, whereas for the substitute \kmer{s}, \ac{SpGEMM} and the formation
of $\mS$ dominate the computation.
The formation of $\mA$ or $\mS$ often takes less time than the \ac{SpGEMM}s at
the smaller node counts.
However, with increasing number of nodes, the percentage of time spent in
\ac{SpGEMM} increases as opposed to that of matrix formation, which indicates
that \ac{SpGEMM} is less scalable.

In Figure~\ref{fig:dissection-single}, we investigate how each component of
\ac{\swname} scales.
The results in the plot at the top are obtained with \texttt{Metaclust50-2.5M}
with exact \kmer{s} and the results in the plot at the bottom are obtained with
the same dataset with 25 substitute \kmer{s}.
In both plots, the bottleneck for scalability seems to be the \ac{SpGEMM}
operations.
Other components either take too short time or scale relatively better.

\begin{figure}[t]
    \centering
    \includegraphics[width=0.90\columnwidth]{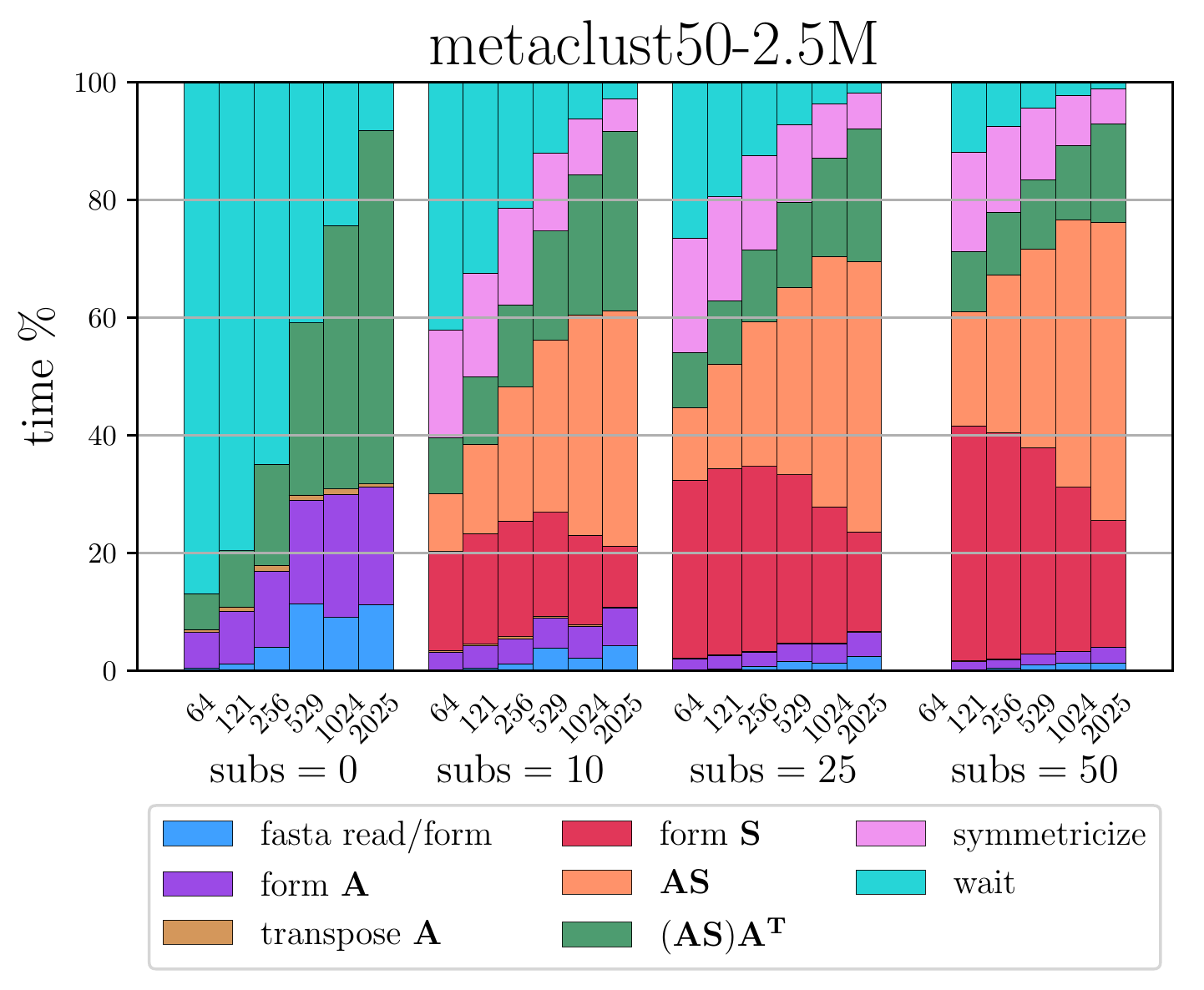}
    \caption{Percentage of time spent in various components of \ac{\swname}
      plotted against increasing number of nodes for different number of
      substitute \kmer{s}.}
    \label{fig:dissection-all}
\end{figure}

\subsection{Precision and Recall}
\label{sec:prec-recall}
In our evaluation, \ac{\swname}, \ac{MMseqs2}, and LAST perform alignment and
generate alignment statistics for sequence pairs to form the similarity graph
$G$.
%
Then, $G$ is clustered with \ac{HipMCL}~\cite{azad2018hipmcl} to discover
possible protein families.
To determine the edge weights in $G$, we use two different similarity measures:
\ac{ANI} and \ac{NS}.
When we use \ac{ANI}, the alignments with \ac{ANI} less than 30\% and shorter
sequence coverage less than 70\% are eliminated from the similarity graph and
the edge weight $w(s_i,s_j)$ is set to the \ac{ANI} of sequences $s_i$ and
$s_j$.

%
In \ac{NS}, we use the raw alignment score normalized with respect to the
shorter sequence.
A motivation for this is that computing \ac{NS} score is cheaper than \ac{ANI}
as the former does not necessitate a trace-back step in the alignment.
Although \ac{NS} may not be as accurate as \ac{ANI}, it still accommodates a
rough information that the clustering algorithm can effectively make use of.
We use a number of different substitute \kmer{s} $\{0,10,25,50\}$ for
\ac{\swname} to measure its effect on the quality of the clusters.
Similarly, we utilize sensitivity values $\{1,5.7,7.5\}$ for \ac{MMseqs2} and
$\{100,200,300\}$ for LAST.
The default for the sensitivity is 5.7 in \ac{MMseqs2}.

\begin{figure}[t]
    \centering
    \includegraphics[width=0.73\columnwidth]{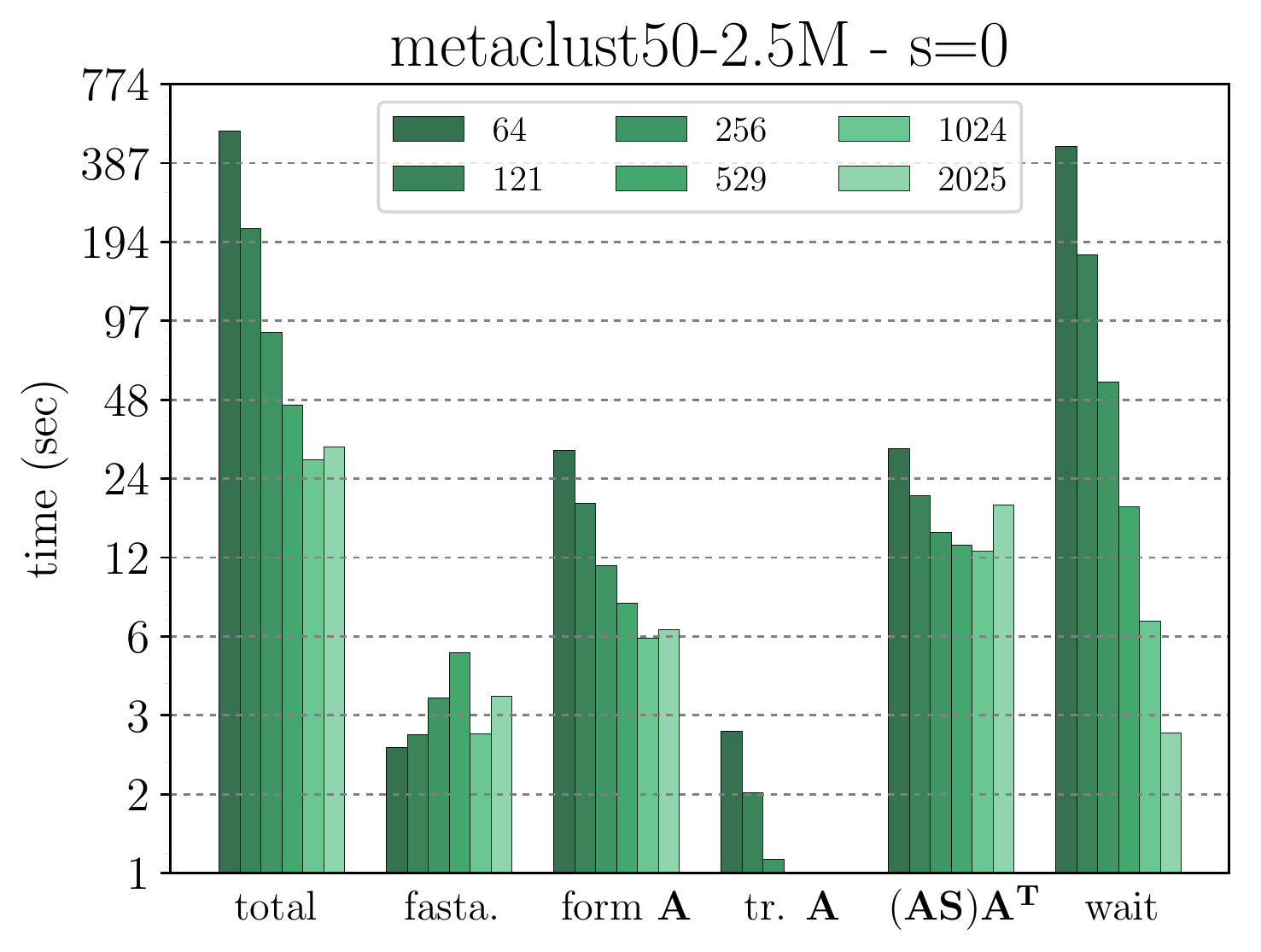}
    \includegraphics[width=0.73\columnwidth]{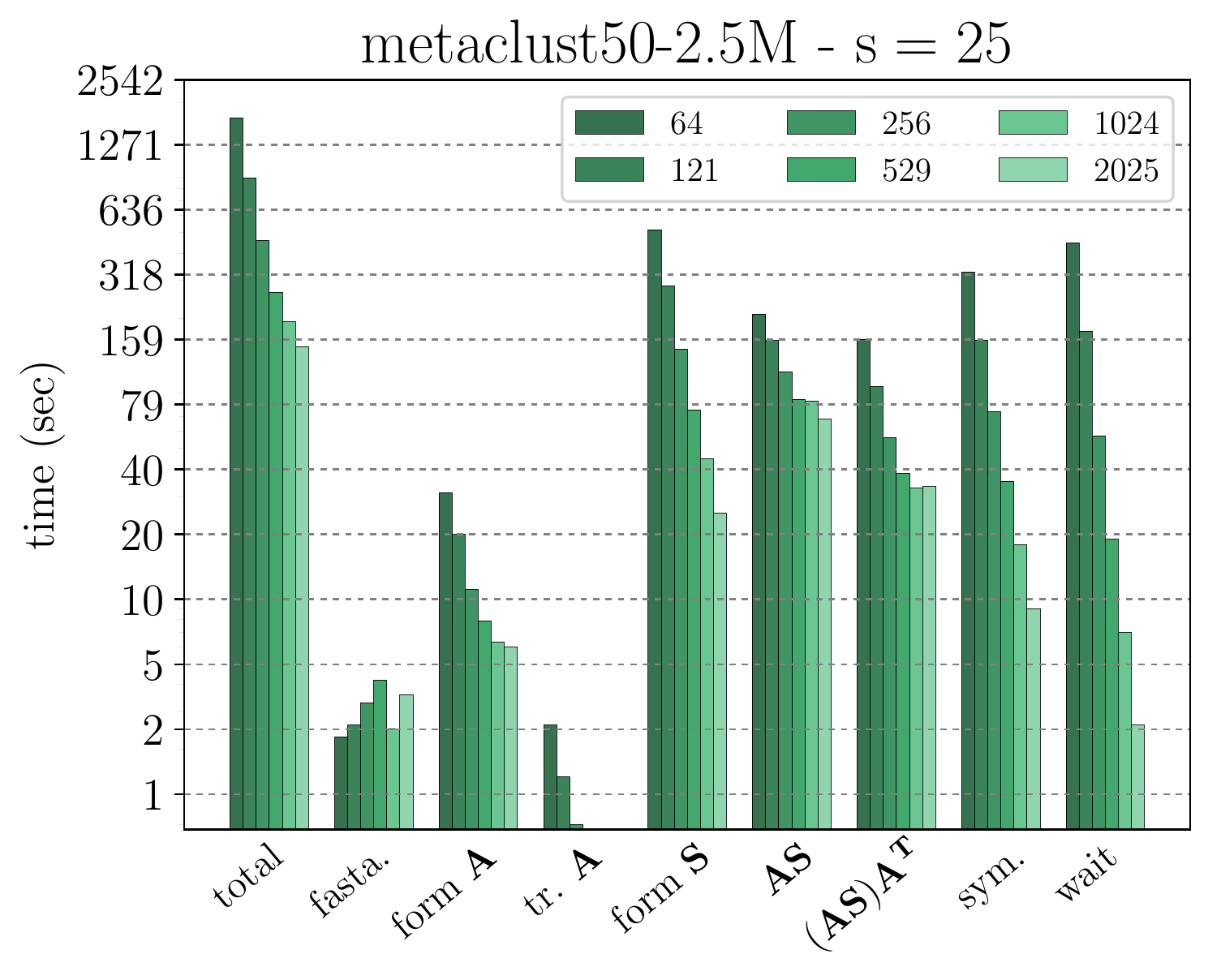}
    \caption{Scaling behavior of various components.}
    \label{fig:dissection-single}
\end{figure}

To compute precision and recall, we use the clusters produced by \ac{HipMCL}
against the protein families in \texttt{SCOPe} dataset.
Particularly, we use weighted precision and recall as defined in the
protein clustering studies~\cite{Bernardes2015}.
The weighted precision metric penalizes clusters that contain proteins from
multiple families while the weighted recall metric penalizes clusters that are
split into multiple families.
Figure~\ref{fig:prec-recall} shows the values obtained by the compared schemes
in terms of weighted precision and recall.
As an important note, the precision range in Figure~\ref{fig:prec-recall} is
between 0.65 and 0.90, and the recall range is between 0.48 and 0.62.

Figure~\ref{fig:prec-recall} shows that one can use different number of
substitute \kmer{s} to adjust the sensitivity, where increasing the number of
substitute \kmer{s} increases recall at the expense of precision.
In the figure, the range in which the precision and recall vary is somewhat
limited.
We note that a more comprehensive spectrum can be obtained by altering the
parameters given to \ac{SeqAn}.
%
%
Except for \ac{\swname}-\ac{SW}-\ac{NS}, \ac{\swname} shows competitive
performance compared to \ac{MMseqs2} and LAST.

Comparing \ac{SW} and \ac{XD} with fixed parameters except for the alignment
mode shows that utilizing \ac{SW} achieves slightly better recall at the expense
of slightly, or sometimes substantially, worse precision.
%
%
A second important point is that \ac{NS} proves to be viable compared to the
\ac{ANI} score.
This is especially the case for \ac{\swname} with \ac{XD} alignment and
\ac{MMseqs2}.
However, \ac{\swname} with \ac{SW} alignment seems to be more sensitive to
\ac{NS}.
Notably, we do not apply any cut-off threshold when using \ac{NS} in the
similarity graph.
Hence, while \ac{XD} may be able to eliminate some of the low-score alignments,
this is not the case for \ac{SW}.
Alignment libraries aside, the clustering algorithm (Markov Clustering in our
case) may also be playing an important role in closing the quality gap between
different weighting schemes used for the edges in the similarity graph.

The common \kmer{} threshold causes a 2\%-3\% loss in recall.
This loss is arguably small when weighted against the benefits it brings in
runtime.
In many cases, applying this threshold results in more than 90\% reduction in
the alignments \ac{\swname} performs and leads to drastic gains in parallel
runtime.

\begin{figure}[t]
    \centering
    \includegraphics[width=0.89\columnwidth]{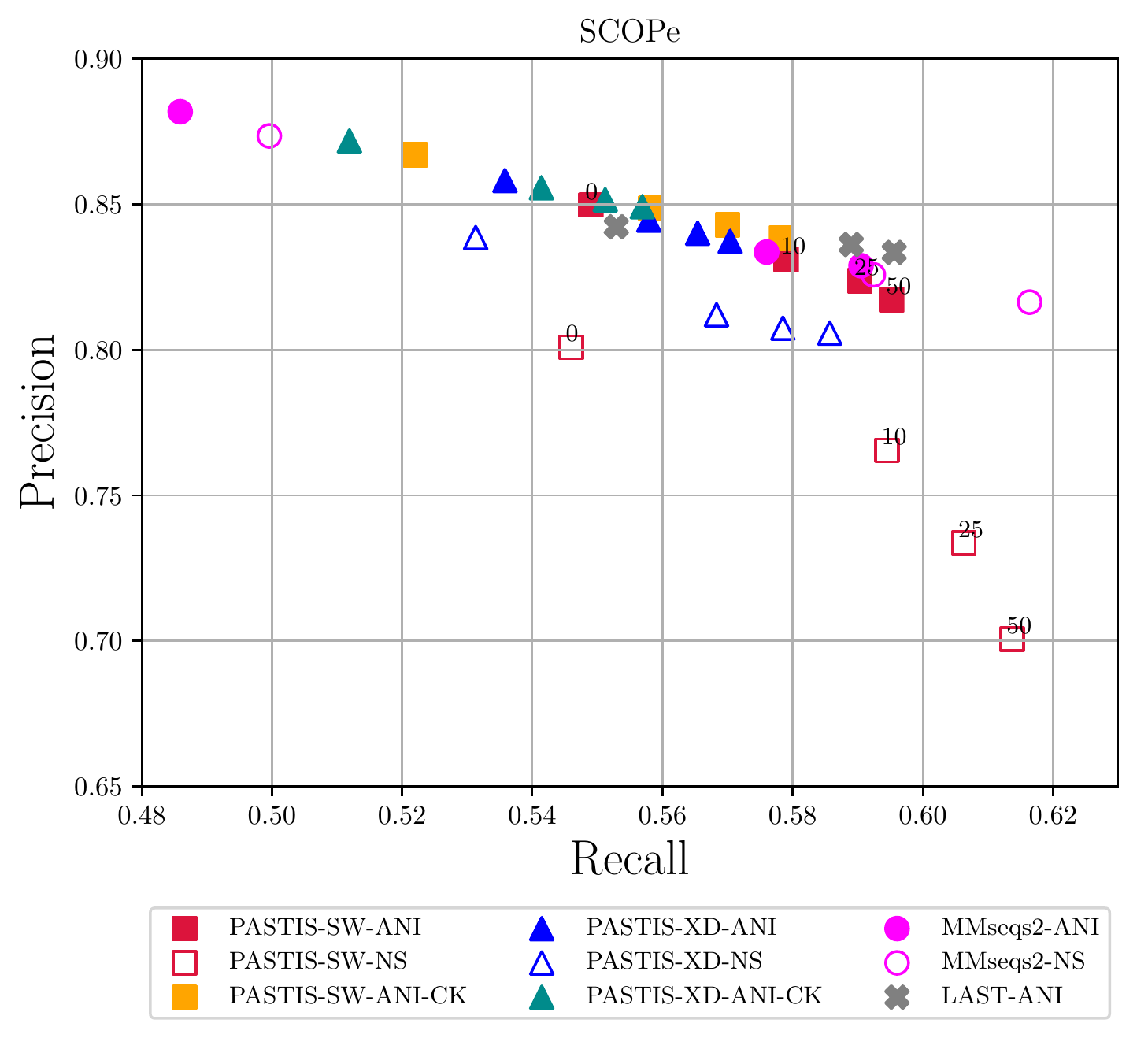}
    \caption{Precision and recall of \ac{\swname}, \ac{MMseqs2}, and LAST for
      various parameters.  The annotation text (over
      \ac{\swname}-\ac{SW}-\ac{ANI} and \ac{\swname}-\ac{SW}-\ac{NS}) indicate
      the number of substitute \kmer{s} utilized in the corresponding setting.
      They are omitted for \ac{\swname}-\ac{XD} variants in order to make the
      figure readable.  The sensitivity parameters for \ac{MMseqs2} variants
      range from left to right as 1 (low), 5.7 (default), 7.5 (high). The
      sensitivity parameters for LAST range from left to right as 100, 300,
      500.}
    \label{fig:prec-recall}
\end{figure}

In Table~\ref{tb:cc}, we investigate the effect of directly using connected
components of the similarity graph as protein families.
Here, the alignments with less than 30\% \ac{ANI} score and less than 70\%
coverage are eliminated.
The connected components might be appealing to avoid running an expensive
clustering algorithm.
Components generated using \ac{MMseqs2} and LAST exhibit higher quality than the
ones generated by \ac{\swname}.
On one hand, using substitute \kmer{s} without clustering causes substantial
precision penalty as the number of connected components gets very small with
increasing number of substitute \kmer{s}.
Therefore, clustering is indispensable when substitute \kmer{s} are used within
\ac{\swname}.
On the other hand, \ac{\swname} with exact \kmer{s} may be a viable option when
clustering cannot be performed.

\section{Conclusion} \label{sec:conclusion}
We developed a scalable approach to protein similarity search within our
software \ac{\swname}.
\ac{\swname} harnesses the expressive power of sparse matrix operations and
those operations' strong parallel performance to enable construction of huge
protein similarity networks on thousands of nodes in parallel.
The substitute \kmer{s} in \ac{\swname} facilitate better recall and enables
controlling sensitivity.
With several optimizations, \ac{\swname} illustrates how linear algebra can
provide a flexible medium to tackle problems in computational biology that deal
with huge datasets and require resources that cannot be met with small- or
medium-scale clusters.
%

Unlocking the true power of distributed-memory supercomputers for protein sequence
similarity search will help accelerate the mining of functional 
and taxonomical diversity of the protein world, especially when analyzing large metagenomic datasets.
Identification of biosynthetic gene clusters~\cite{weber2015antismash}, discovery of novel protein families~\cite{sberro2019large}, uncovering the world of viruses~\cite{schulz2020giant}, and routine functional~\cite{franzosa2018species}
and taxonomical~\cite{menzel2016fast} classification
from large metagenomic samples all depend on the ability to perform protein sequence similarity
search at scale.  \ac{\swname} is a step towards enabling new biological discoveries at extreme scale
using the fastest supercomputers. One of the next major milestones in \ac{\swname} development
is to take advantage of accelerators such as GPUs in distributed-memory 
protein sequence similarity search.

As future work, we first plan to address the issues related to memory
requirements of \ac{\swname} as it occasionally runs out of memory at small node
counts.
A direction in this regard is the partial formation of the output matrix and
once this partial information is obtained to run the alignment and free the
corresponding memory.
Another future avenue is to perform an analysis of \kmer{s} in a pre-processing
stage to see whether some of them can be eliminated without sacrificing
recall too much.
We also plan to conduct an performance analysis of enhanced pipeline with
clustering.

\begin{table}[t]
  \centering
  \caption{Connected components as protein families.}
  \hspace*{0.6cm}
  \scalebox{0.83} {
    \begin{tabular}{l l r r r r}
      \toprule
      & & \multicolumn{4}{c}{Number of substitute \kmer{s}} \\
      \cmidrule(r{4pt}){3-6}
       &  & 0 & 10 & 25 & 50  \\
      \midrule
      \multirow{2}{*}{\ac{\swname}-\ac{SW}} & precision & 0.67 & 0.38 & 0.28 & 0.22   \\
                                            & recall    & 0.67 & 0.77 & 0.83 & 0.87   \\
      \midrule
      \multirow{2}{*}{\ac{\swname}-\ac{XD}} & precision & 0.69 & 0.55 & 0.46 & 0.39 \\
      & recall    & 0.64 & 0.69 & 0.73 & 0.76 \\
      \bottomrule
    \end{tabular}
  }
  \newline
  \vspace{0.5em}
  \newline
  \hspace*{1.0cm}
   \scalebox{0.83} {
     \begin{tabular}{l l r r r}
      & & \multicolumn{3}{c}{Sensitivity} \\
      \cmidrule(r{4pt}){3-5}
        &  & 1 & 5.7 & 7.5 \\
        \midrule
        \multirow{2}{*}{\ac{MMseqs2}} & precision & 0.77 & 0.75 & 0.75 \\
                                      & recall    & 0.60 & 0.71 & 0.72 \\
        \bottomrule
    \end{tabular}
    }
   \newline
  \vspace{0.5em}
  \newline
  \hspace*{0.3cm}
  \scalebox{0.83} {
    \begin{tabular}{l l r r r}
      & & \multicolumn{3}{c}{Max initial matches} \\
      \cmidrule(r{4pt}){3-5}
        &  & 100 & 200 & 300 \\
        \midrule
        \multirow{2}{*}{LAST} & precision & 0.76 & 0.76 & 0.76 \\
                              & recall    & 0.68 & 0.70 & 0.70 \\
        \bottomrule
    \end{tabular}
    }

  \label{tb:cc}
\end{table}

\section*{Acknowledgments}
This work used resources of the NERSC supported by the Office of Science of the
DOE under Contract No. DEAC02-05CH11231.

This work is supported in part by the Advanced Scientific Computing Research (ASCR) Program of the Department of Energy Office of Science under contract
No. DE-AC02-05CH11231, and in part by the Exascale Computing Project
(17-SC-20-SC), a collaborative effort of the U.S. Department
of Energy Office of Science and the National Nuclear Security
Administration.

Georgios A. Pavlopoulos was supported by the Hellenic Foundation for Research
and Innovation (H.F.R.I) under the ``First Call for H.F.R.I Research Projects to
support Faculty members and Researchers and the procurement of high-cost
research equipment grant'', GrantID: 1855-BOLOGNA.

\bibliographystyle{IEEEtran}
\bibliography{references}

\end{document}